\documentclass{aa}
\usepackage[varg]{txfonts}
\usepackage{amsmath,amsfonts,amssymb}
\usepackage{graphicx}
\usepackage[colorlinks=true, allcolors=blue]{hyperref}
\usepackage{gensymb}
\usepackage{mathtools}  
\usepackage{braket}     
\bibpunct{(}{)}{;}{a}{}{,} 
\usepackage[]{natbib} 

\begin{document}
%
\title{Frequency chirped continuous-wave sodium laser guide stars}
\author{
F. Pedreros Bustos\inst{1} 
\and R. Holzl\"ohner\inst{2} 
\and S. Rochester\inst{3} 
\and D. Bonaccini Calia\inst{2} 
\and J. Hellemeier\inst{2,4} 
\and D. Budker\inst{1,5}
}

%
%
\institute{
Johannes Gutenberg University, Helmholtz Institute Mainz, Staudingerweg 18, 55128 Mainz, Germany \\
\email{[pedreros;budker]@uni-mainz.de}
\and European Southern Observatory, Karl-Schwarzschild-Str. 2, 85748 Garching b. M\"unchen, Germany \\
\email{[rholzloe;dbonacci;jhelleme]@eso.org}
\and Rochester Scientific LLC, 2041 Tapscott Ave., El Cerrito 94530 CA, USA \\
\email{simon@rochesterscientific.com}
\and University of British Columbia, 6224 Agricultural Road, Vancouver, BC V6T1Z1, Canada 
\and Department of Physics, University of California Berkeley, Berkeley, CA 94720-7300, USA}
\date{Received: date / Revised version: date}
%
\abstract{
We numerically study a method to increase the photon return flux of continuous-wave laser guide stars using one-dimensional atomic cooling principles. The method relies on chirping the laser towards higher frequencies following the change in  velocity of sodium atoms due to recoil, which raises atomic populations available for laser excitation within the Doppler distribution. The efficiency of this effect grows with the average number of atomic excitations between two atomic collisions in the mesosphere. We find the parameters for maximizing the return flux and evaluate the performance of chirping for operation at La Palma. According to our simulations, the optimal chirp rate lies between 0.8--1.0~MHz/$\mu$s and an increase in the fluorescence of the sodium guide star up to 60\% can be achieved with current 20 W-class guide star lasers.
} 
\keywords{instrumentation: adaptive optics -- methods: numerical -- atomic processes -- recoil}

\maketitle

\section{Introduction}
\label{intro}

Laser guide stars (LGS) along with adaptive optics (AO) systems are fundamental tools for modern observatories, since together they allow to reach the telescope's diffraction limit. Three new large-telescope projects, the Giant Magellan Telescope (GMT), Thirty Meter Telescope (TMT) and Extremely Large Telescope (ELT) will employ several LGS units as they play a significant role in reaching the ultimate optical performance of their instruments. 

Sodium LGS exploit the naturally occurring atomic layer in the upper mesosphere between 85 km and 100 km altitude. Laser light tuned to the wavelength of 589.1591~nm in vacuum, resonant with the $3^2$S$_{1/2} - 3^2$P$_{3/2}$ transition of sodium (also known as D$_2$ line), is absorbed and spontaneous emission from the sodium layer is generated. The light from the LGS is used as a reference for the AO system on the telescope to compensate the distortions introduced by the atmospheric turbulence in the wavefront of an astronomical object. A brighter LGS can reduce the residual error and increase the performance of an AO system, which is particularly important for observations in the visible part of the spectrum and during daytime.

Over the last decades, several laser formats have been explored to create brighter sodium LGS~\citep{dOrgeville:2016}. Single-frequency continuous-wave (CW) lasers based on Raman-fiber-amplifier technology are currently in use because they can efficiently pump the sodium layer generating high photon-return flux and offer a high reliability for operation in astronomical observatories~\citep{Bonaccini:2014}. In spite of the high efficiency of current LGS technology, there are three factors that reduce the return from an LGS, namely Larmor precession, transition saturation, and recoil~\citep{Holz:2010a}.     

At large angles between the laser beam and the geomagnetic field lines in the sodium layer, Larmor precession redistributes atomic populations among the ground-state magnetic sublevels, which reduces the number of atoms that can be optically pumped on the stronger optical transition. As a consequence, fluorescence from sodium atoms is reduced. Synchronous pumping of the sodium layer with intensity or polarization modulation of the laser beam has been proposed and demonstrated to mitigate this effect~\citep{Higbie:2011,Kane:2018,Pedreros:2018b}. An additional laser (or spectral line) with a frequency offset of +1713~MHz from the D$_2$ line can be used as a ``repumper'', in order to compensate the effect of downpumping of atomic populations from the $F=2$ to $F=1$ ground state (where $F$ is the total angular momentum), which becomes severe as the effect of Larmor precession takes place. 

When the irradiance in the sodium layer approaches the saturation intensity, there is an increasing probability of stimulated emission by an excited atom. The stimulated emitted photon is directed into space and, as a result, the net photon-return flux seen by an observer on earth is reduced. Finally, at larger irradiances the atomic medium becomes more transparent to resonant light and the LGS is less efficient. 

\begin{figure}[t]
\begin{center}  
   \includegraphics[width=0.8\linewidth]{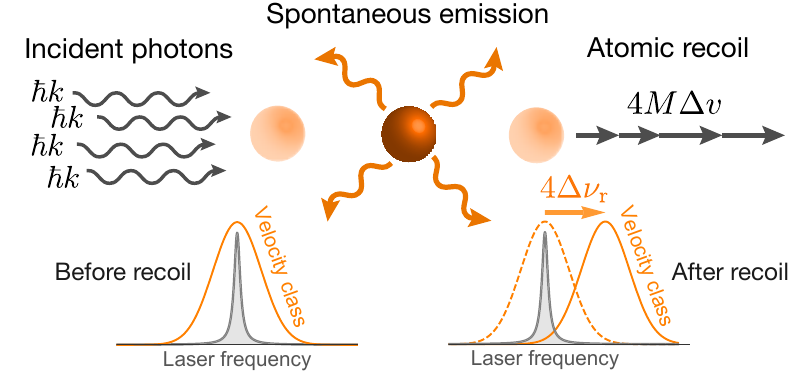} 
    \caption{Upon spontaneous emission, an atom recoils in the same direction of the incoming photons due to linear momentum conservation. On each absorption/emission cycle, the resonance frequency of the recoiled atom is raised by $\Delta \nu_\text{r}$ with respect to the frequency of the laser. After many cycles, recoil can shift the velocity class out of optical resonance.}  \label{fig:chirping}
\end{center}
\end{figure}

In this paper we address the third factor that reduces the LGS photon-return flux, i.e. recoil. When a sodium atom absorbs and reemits a photon, there is a small change of its velocity due to conservation of momentum given by:

\begin{equation}
\Delta v = \hbar k/M,
\end{equation}
where $\hbar$ is the reduced Planck constant $\hbar = h/2\pi = 1.05\times10^{-34}$~J$\cdot$s, $k=2\pi/\lambda$ is the wavenumber, $\lambda$ is the resonant wavelength of sodium, and $M=3.81\times 10^{-26}$~kg is the atomic mass of sodium. In the mesosphere, spontaneously emitted photons are radiated randomly in all directions, although with a spatial point symmetry about the center of the atom when the medium is excited with polarized light. Therefore, the average recoil due to spontaneous emission is zero. However, as the absorbed laser photons (each with momentum $\hbar k$) travel in the same direction, there is a net increase in the velocity of the atomic ensemble in the direction of the incoming photons as shown in Fig.~\ref{fig:chirping}. This net recoil translates into a Doppler shift equal to $\Delta \nu_\text{r} = \Delta v /\lambda = 50.2$~kHz, which changes the resonant frequency of the atom with respect to the incoming light (redshift of the laser light as seen from the moving atom, blueshift as seen from the emitter when tracking the resonance frequency). After many absorption/emission cycles, an atom accumulates a frequency shift until it leaves the laser beam (transit time) or until it collides with another atom or molecule. The number of cycles can be estimated as:
\begin{equation}
n_c = T_t/T_c,
\end{equation}
where $T_t$ is the transit time (or the collision coherence time) and $T_c$ is the cycling time. At an irradiance of 62.6~W/m$^2$ (saturation intensity of sodium) the cycling time, hence the mean time between two spontaneous emissions is $T_c=64$~ns for the case of a pure two-level system and circularly polarized light. Given an average recoil frequency shift of $\Delta \nu_\text{r} = 50.2$~kHz per cycle, the rate of change of the atom's resonant frequency is $(50.2~\text{kHz})/(64~\text{ns})=0.78$~MHz/$\mu$s. For the purpose of illustration, assuming a typical mean time between collisions in the mesosphere of 35~$\mu$s, it takes approximately $546$ cycles on average before the atomic velocities are randomized due to collisions with another particle. Then, the accumulated frequency shift of one atom during the collision coherence time is $\delta \nu = n_c \Delta \nu_\text{r} = 28$~MHz. However, after only $\approx$~200 cycles the atom’s resonant frequency shifts by 10~MHz which corresponds to the natural linewidth of sodium. Beyond this point, the atom is no longer resonant with a single-frequency light source. Nevertheless, after several collisions the atom may re-enter the resonant velocity class and become available for optical pumping again. In this simplified discussion we have ignored other relaxation effects such as downpumping and Larmor precession, but it shows, in principle, how recoil decreases the efficiency of optical excitation of sodium. 

One way to deal with the effect of recoil is to compensate the atomic redshift by blueshifting the optical frequency of the excitation laser. This technique, generally called frequency chirping, was first proposed for atomic cooling~\citep{Hansch:1975,Wineland:1978}, and later demonstrated as a mechanism to slow down the velocity of an atomic beam~\citep{Ertmer:1985,Watts:1986,Littler:1991,Phillips:1998} and for trapping neutral atoms in a confined volume with sub-Kelvin temperatures~\citep{Ashkin:1978,Chu:1985}. Frequency chirping for increasing the brightness of an LGS was proposed by \citet{Jeys:1992} and first numerically modeled in pulse-train excitation of sodium by \citet{Bradley:1992}. These simulations showed a significant increase in steady-state upper-level populations by a factor of two compared to an unmodulated laser. An alternative model of chirped LGS was presented by \citet{Kibblewhite:2008} using a Monte Carlo rate-equation approach for pulsed and CW excitation formats, including the effects of optical pumping, downpumping, recoil, and magnetic fields. Measurements carried out by the same group using the Chicago/Palomar sum-frequency micro/macro pulsed laser system ($1$~ns pulses at 100~MHz under a 150~$\mu$s pulse envelope at 400~Hz) showed an enhancement of the return flux by a factor of 1.8 using laser chirping~\citep{Kibblewhite:website}. Additionally, \citet{Hillman:2008} presented another model based on rate equations for the CW laser format. Sophisticated on-sky experiments using two overlapping CW lasers (one pump laser and a second probe laser) by the same group attempted to measure the atomic velocity distribution of sodium to show the effect of recoil in the mesosphere; however, the experiments yielded inconclusive results.

Here, we present a study of the return flux of an LGS excited with a single-frequency CW laser and with frequency chirping, using a state-of-the-art numerical model based on optical Bloch equations. This work is, to the best of our knowledge, the most in-detailed investigation to quantify the benefits of chirping with single-frequency CW lasers and to determine the best possible chirp rate. Section~\ref{sec:method} describes the model and the parameters used for the calculations. Results are presented in Sect.~\ref{sec:results} showing the changes in return flux due to varying a set of parameters. Finally, we discuss the optimal chirping parameters and the benefits and possible limitations of the chirping scheme for laser guide stars.  

\section{Method} \label{sec:method}

Laser excitation of mesospheric sodium and the expected fluorescence are modeled using the optical Bloch equations in the atomic density matrix formalism~\citep{Rochester:2012}. The density matrix describes the statistical state of an ensemble of atoms in the 24 Zeeman sublevels of the state space of the Na D$_2$ transition. As the transition is Doppler broadened the density matrix is a function of the atomic velocity along the laser propagation direction. This velocity dependence is treated by dividing the Doppler distribution into discrete velocity groups ($\Delta v_\text{v.g.}$). The evolution of the density matrix is given by the generalization of the Schr\"odinger equation (the Liouville-von Neumann equation):
\begin{equation}
\dfrac{d}{dt}\rho = \frac{1}{i\hbar}[H,\rho] + \Lambda(\rho) + \beta, \label{eq:schrodinger}
\end{equation}

\noindent where the atomic level structure and interaction with external fields are described by the total Hamiltonian $H$. For chirped light, $H$ depends on time due to the frequency sweep. The term $\Lambda$ accounts for relaxation and repopulation processes, namely spontaneous decay, collisions, changes in atomic velocity and exit of atoms from the light beam;  $\beta$ describes the additional repopulation process due to the entrance of atoms into the beam, which is independent of $\rho$. Recoil is modeled phenomenologically by allowing a fraction $\Delta v/\Delta v_\text{v.g.}$ of the excited-state atoms in each velocity group to be transferred upon decay into the next higher velocity group. 

Equation \ref{eq:schrodinger} provides a linear system of differential equations for the density-matrix elements (Bloch equations), which can be written as $\dot{\rho} = A \rho + b$, where $A$ and $b$ are a matrix and a vector, respectively, and $\rho$ is a column vector of $n_\text{v.g.}\times24^2$ density-matrix elements (where $n_\text{v.g.}$ is the total number of velocity groups).     

The density-matrix evolution equations are generated using the LGSBloch package for Mathematica, which is based on the Atomic Density Matrix package\footnote{Available at \url{http://rochesterscientific.com/ADM/}}. The system of ordinary differential equations (ODE) is solved using code based on the open-source ODE solver CVODE from the SUNDIALS package \citep{Sundials}. 

The fluorescent photon flux per solid angle emmitted in a given direction can be found from the solution for $\rho$ as the expectation value of a fluorescence operator~\citep{Corney:1977,Auzinsh:2010}.

\subsection{Standard parameters}
 
Besides fundamental constants, the model presented above needs several input parameters related to the laser beam characteristics and the kinematics of gases in the mesosphere that must be estimated beforehand. The accuracy of these input parameters plays a fundamental role in the prediction of the return flux of an LGS. 

In order to model the laser beam size, we assume an effective Gaussian beam in the mesosphere whose full-width-half-maximum (FWHM) has been estimated using simulated physical optics propagation through a turbulent atmosphere with a seeing of 1.0 arcsec at 500 nm and zenith~\citep{Holz:2008}. 

\begin{figure}[t]
\begin{center}  
   \includegraphics[width=0.99\linewidth]{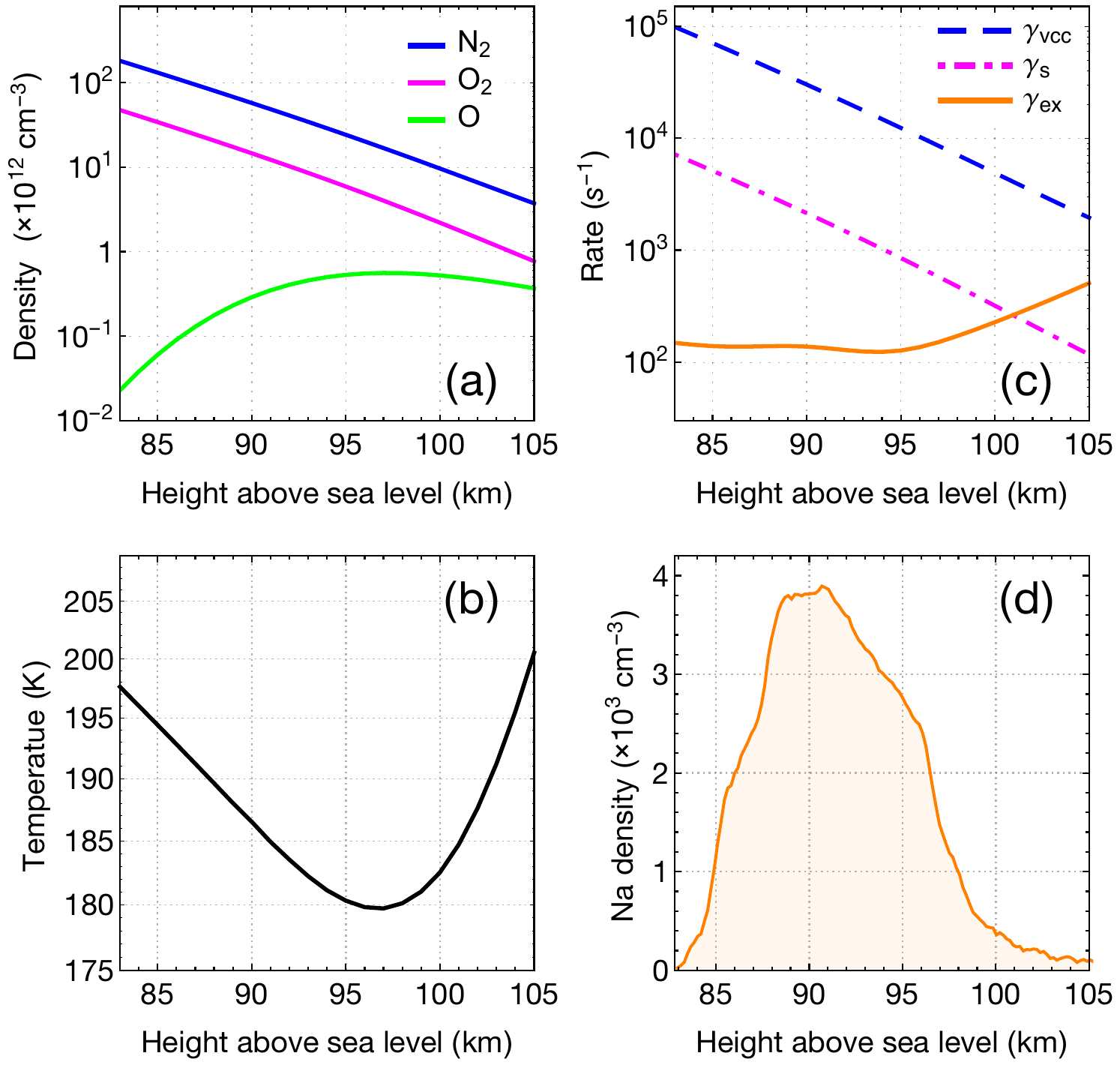} 
    \caption{Vertical profiles of (a) molecular number densities, (b) temperature, and (c) collisions and beam exchange rate ($\gamma_\text{vcc}:$ velocity-changing collisions, $\gamma_\text{s}:$ spin-damping collisions, $\gamma_\text{ex}:$ atom-beam exchange) for La Palma. (d) Reference vertical sodium density profile obtained with the Large Zenith Telescope LIDAR facility in Vancouver, (BC, Canada), normalized with a sodium column abundance of $C_\text{Na}=4.0 \times 10^{9}$~atoms$/\text{cm}^2$.}  \label{fig:relaxation}
\end{center}
\end{figure}

Vertical profiles of atmospheric temperatures, molecular number densities, and winds were used as inputs to estimate atomic collisions and transit times. Profiles of molecular number densities and temperature in the mesosphere are obtained from the MSISE-00 atmospheric model\footnote{\url{https://ccmc.gsfc.nasa.gov/modelweb/models/nrlmsise00.php}}~\citep{MSISE:2010}, while wind profiles are obtained from the Horizontal Wind Model 2014 (HWM14)~\citep{HWM14} assuming a quiet atmosphere. For example, Figs.~\ref{fig:relaxation}.(a) and \ref{fig:relaxation}.(b) show the relevant molecular densities and temperature profiles between 83~km and 105~km above sea level at La Palma (latitude:$+28.75$\degree, longitude:$-17.89$\degree, elevation: 2396 meters above sea level) used as inputs to calculate the rate of atomic collisions in the mesosphere.
%

The diffusion coefficient of sodium in the gas mixture can be estimated from Chapman-Enskog gaseous diffusion theory~\citep{Chapman:1970} and from the diffusion model in multicomponent systems by \citet{Fairbanks:1950}. Along with wind profiles from the HWM14 model and calculations of the beam size in the mesosphere, we can estimate the exchange rate of sodium atoms in the laser beam. 

In a mixture of gases, binary collision rates between a particle of mass $M_1$ and particles of mass $M_2$ and number density $n_2$ can be calculated with
\begin{equation}
\gamma_{12} = n_2 \sigma_{12}\sqrt{\frac{8 k_\text{B}T}{\pi} \left( \frac{1}{M_1} + \frac{1}{M_2}\right)},
\end{equation}
where $k_\text{B}=1.38\times10^{-23}$~J/K is the Boltzmann constant, and $\sigma_{12}=\pi(r_1 +r_2)^2$ is the collisional cross section with the Van-der-Waals radii of each particle $r_1$ and $r_2$. We follow the approach for calculating atomic collision rates described by \citet{Holz:2010a} using updated versions of the atmospheric models. Figure~\ref{fig:relaxation}.(c) shows the atomic collision rates above La Palma following the presented model. The dominant relaxation mechanism is the velocity-changing collisions due to Na$-$O$_2$ and Na$-$N$_2$ colliding atoms.


A standard vertical sodium profile obtained with a lidar at the Large Zenith Telescope (LZT)~\citep{Pfrommer:2014} is shown in Fig.~\ref{fig:relaxation}.(d). This profile is used throughout the simulations as a reference for calculating the final fluorescence from the sodium layer. Although the sodium profile is highly variable, we believe that a measured typical profile yields a more realistic modeling of the system than a standard Gaussian profile.

Typically, CW guidestar lasers use an additional spectral line with a frequency offset of $+1713$~MHz relative to the
$3^2$S$_{1/2}(F=2) - 3^2$P$_{3/2}(F'=3)$ transition (D$_2$a line). This sideband pumps atoms from the $F=1$ ground state to the $F'=2$ upper state (D$_2$b line) that  decay into the $F=2$ ground state, and as a result, a high degree of optical pumping can be obtained increasing the brightness of an LGS in a factor of 1.5--2~\citep{Holz:2016}. The additional spectral line is called the repumper. The repumping fraction $q$ is defined here as the fraction of laser power in the repumper line ($P_\textbf{D2b}$) with respect to the power content in the D$_2$a and D$_2$b lines ($P_\textbf{D2a}+P_\textbf{D2b}$), and it is given by
\begin{equation}
q = \dfrac{P_\textbf{D2b}}{P_\textbf{D2a} + P_\textbf{D2b}}. \label{eq:repumping}
\end{equation}  

A summary of the nominal standard parameters used for the simulations is presented in Table~\ref{Table:parameters}. 

\begin{table}[h]
\caption{Nominal parameters used for simulations.}
\label{Table:parameters}
\centering
\begin{small}
\begin{tabular}{lc}
\hline
\hline
\multicolumn{1}{c}{Parameter} & Value \\ \hline
D$_2$a transition wavelength in vacuum ($\lambda$) & 589.15905 nm \\ 
Laser linewidth & 0 MHz \\ 
Polarization & Circular \\ 
Repumping fraction (\textit{q}) & 0.1 \\ 
Repumping frequency offset & $+$1713~MHz \\ 
Sodium column density ($C_\text{Na}$) & $4.0 \times 10^{9}$ atoms/cm$^2$ \\ 
One-way atmospheric transmission at zenith & 0.89 \\
Temperature at 91 km above sea level  & 185 K \\  
Magnetic polar angle ($\theta_B$) & 0\degree \\ 
Beam atom exchange rate ($\gamma_\text{ex}$) & 1/(6 ms) \\ 
Velocity-changing collision rate ($\gamma_\text{vcc}$) & 1/(35 $\mu$s) \\ 
Spin-exchange collision rate ($\gamma_\text{s}$) & 1/(490 $\mu$s) \\ \hline
\end{tabular}
\end{small}
\end{table}

%
%
%

\section{Results} \label{sec:results}

\subsection{Atomic populations and velocity distribution}

\begin{figure}[b]
\begin{center}  
   \includegraphics[width=0.8\linewidth]{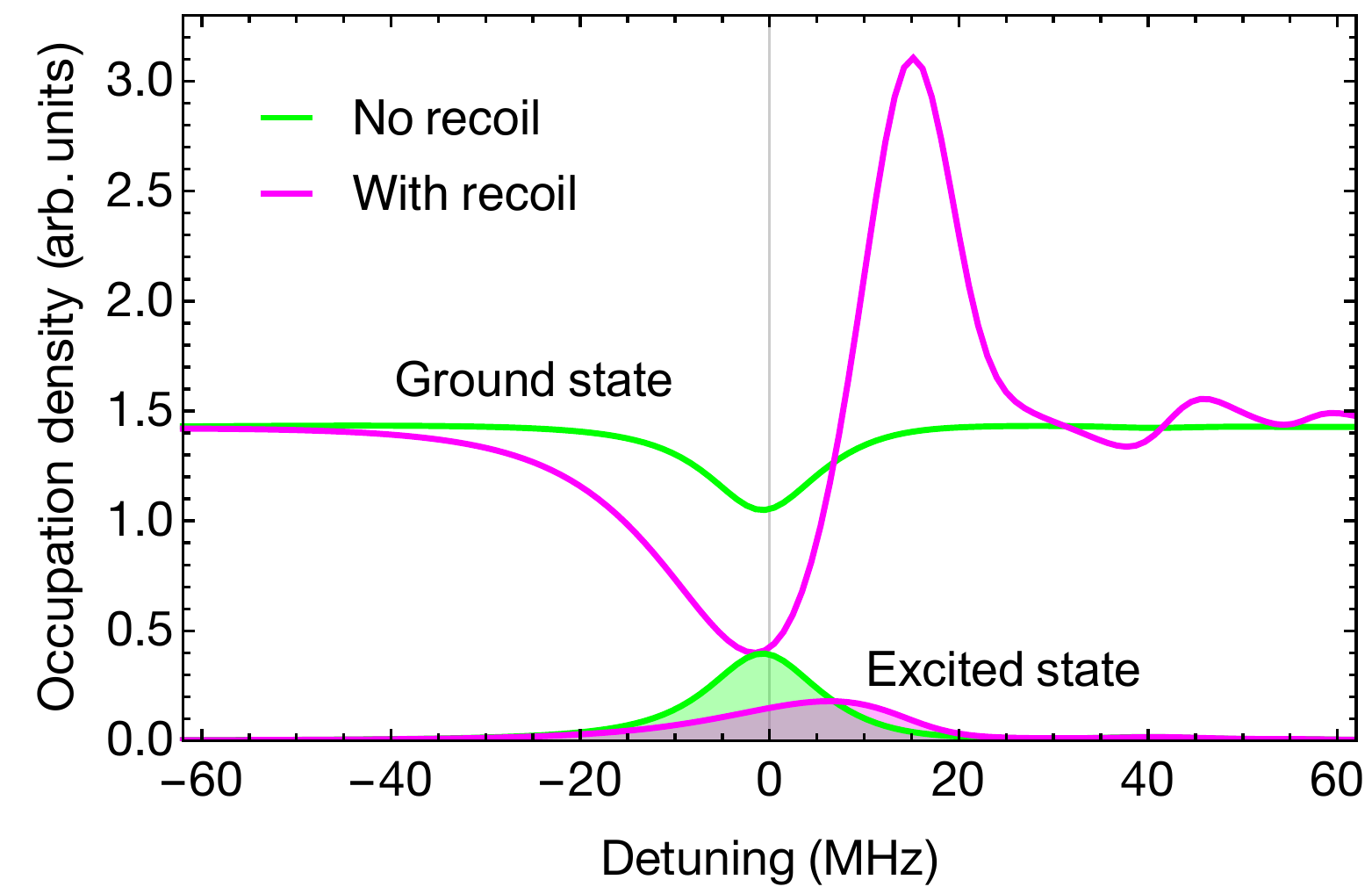} 
    \caption{Comparison of ground-state and excited-state velocity distributions when recoil is included and neglected  in the model for irradiance of 100~W/m$^2$. The additional feature seen at $+45$~MHz in the ground-state distribution (red line) is due to the repumping transition.}  \label{fig:velocity_noRecoil}
\end{center}
\end{figure}

\begin{figure}[t]
\begin{center}  
   \includegraphics[width=0.75\linewidth]{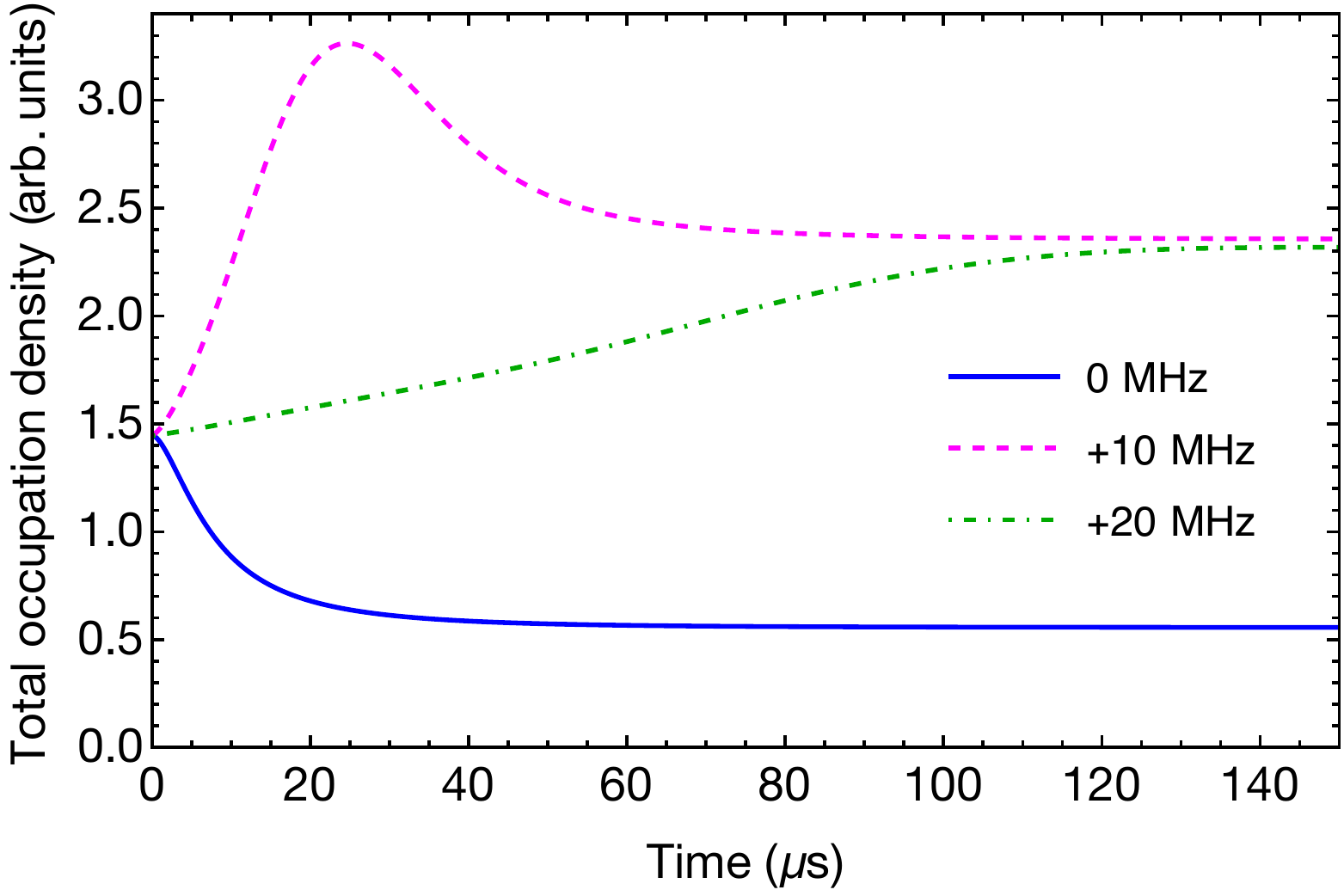} 
    \caption{Ground-state occupation density evolution at $0$~MHz, $+10$~MHz and $+20$~MHz velocity classes with fixed laser frequency for $I=100$~W/m$^2$ and standard parameters, including recoil.}  \label{fig:occupation_density}
\end{center}
\end{figure}

\begin{figure}[b]
\begin{center}  
   \includegraphics[width=0.8\linewidth]{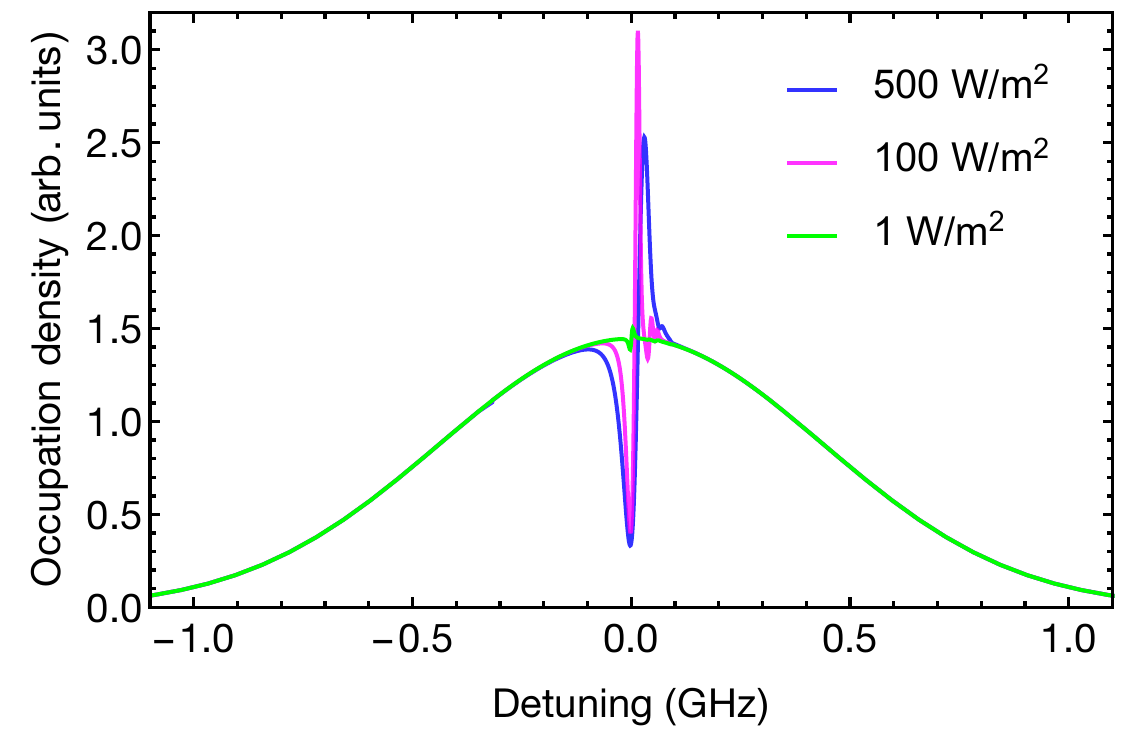} 
    \caption{Hole burning in the Maxwell-Boltzman velocity distribution for light intensities of 1, 100 and 500~W/m$^2$, including recoil and for zero magnetic field.}  \label{fig:velocity_Maxwell}
\end{center}
\end{figure}

\begin{figure*}[h]
\begin{center}  
   \includegraphics[width=0.8\linewidth]{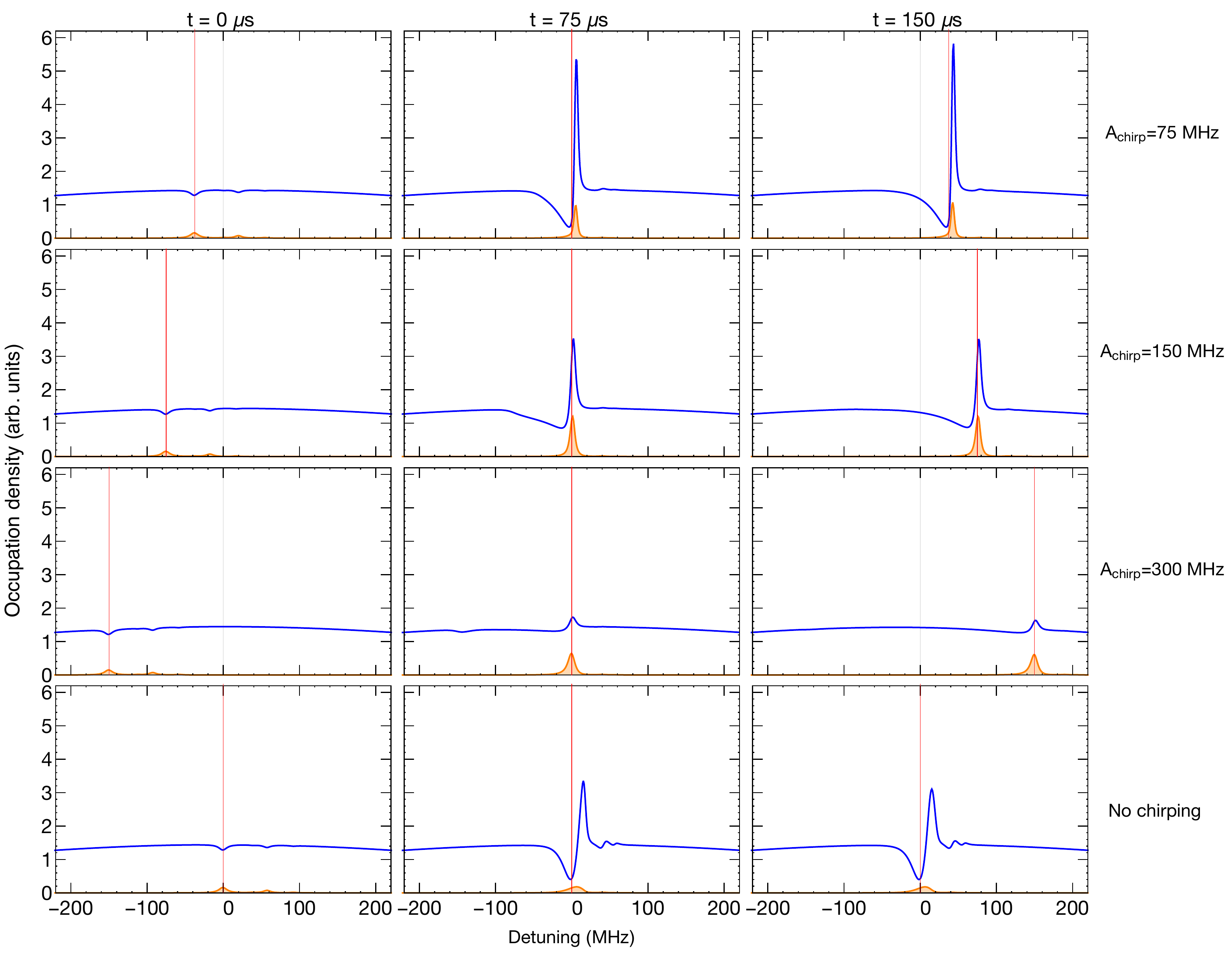} 
    \caption{Time evolution of the velocity distribution around its maximum for irradiance of 100~W/m$^2$ and chirping time $\tau_\text{chirp}=150$~$\mu$s. Ground-state population is indicated in blue and contribution to fluorescence is indicated in orange. The red-vertical line indicates the current detuning in the optical frequency of excitation.}  \label{fig:velocity_sequence}
\end{center}
\end{figure*}

We assume mesospheric sodium in thermal equilibrium, therefore atoms move in all directions with different velocities following the Maxwell-Boltzmann distribution. The most probable speed $v_p$ of sodium atoms at temperature $T$ is given by $v_p = \sqrt{2 k_\text{B}T/M}$, which for $T=185$~K corresponds to $v_p=367$~m/s. When high-intensity single-frequency light propagating in the $z$ direction is used to pump the atomic medium, a depletion of atomic populations in the ground state of the resonant velocity class occurs as a result of saturation of the transition. If one observes the velocity distribution, for instance, via velocity-selective saturation spectroscopy, a hole will be seen at $v_z = (\omega - \omega_0)/k$, where $\omega$ is the optical frequency around the resonance frequency for stationary atoms $\omega_0$, and $k$ is the wavenumber. In the absence of recoil, only a single dip with saturated width $\gamma_s=\gamma \sqrt{1+S_0}$ will be created, where $\gamma$ is the homogeneous width of the transition and $S_0 \geq 0$ is the saturation parameter defined as the ratio of pumping rate to the average relaxation rate at $\omega_0$. However, when recoil shift $\Delta \nu_\text{r}$ is considered in the model, an excess of atomic populations or a bump in the velocity distribution is produced as depicted in Fig.~\ref{fig:velocity_noRecoil}. The peak represents a larger amount of ground-state populations that recoil with increasing velocity towards the neighbor velocity group and that see the laser light with a lowered frequency. On the other hand, atomic population in the excited states diminishes with respect to a system without recoil, because a large fraction of the resonant atoms have shifted out of resonance and cannot be excited any longer. Consequently, atomic recoil reduces return flux for atomic excitation with narrow-band lasers. This result shows that our model properly represents the existing physical mechanisms, and that recoil is an important factor that reduces the efficiency of an LGS.

The time evolution of the ground-state populations at zero detuning, $+10$~MHz, and $+20$~MHz is shown in Fig.~\ref{fig:occupation_density}. In the first 30~$\mu$s, more than half of the atoms resonant with the laser have recoiled, while the population in the neighboring non-resonant velocity class 10~MHz higher has doubled. The effect of recoil continues reducing the population at $+10$~MHz detuning and populating a velocity class at $+20$~MHz from resonance. After 150~$\mu$s, the system reaches a steady state and a large fraction of resonant atoms are lost. The complete velocity distribution at this point is the one shown in Fig.~\ref{fig:velocity_noRecoil}.

The effect of irradiance on the velocity distributions can be observed in Fig.~\ref{fig:velocity_Maxwell}. At low irradiance ($I=1$~W/m$^2$, $S_0\approx 0.02$), the small perturbation at zero detuning is hardly visible. In this regime, the effect of recoil is negligible, as previously concluded by \citet{Milonni:1999}. At irradiances near/above saturation ($I=100$~W/m$^2$, $S_0\approx 1.6$), the hole burning in the resonant velocity class is evident and a large fraction of atoms ($\approx 60$\%) are displaced to the next higher velocity class. With even larger irradiance ($I=500$~W/m$^2$, $S_0\approx 8.0$), the depth of the hole is as large as in the previous case and power broadening is noticeable. The peak of recoiled atoms is lower with respect to the peak at $I=100$~W/m$^2$ as populations are spread over a wider range of the distribution. We believe that our model is able to predict fundamental effects and that the underlying atomic physics mechanisms are being modeled accurately in our simulations.

\begin{figure}[b]
\begin{center}  
   \includegraphics[width=0.8\linewidth]{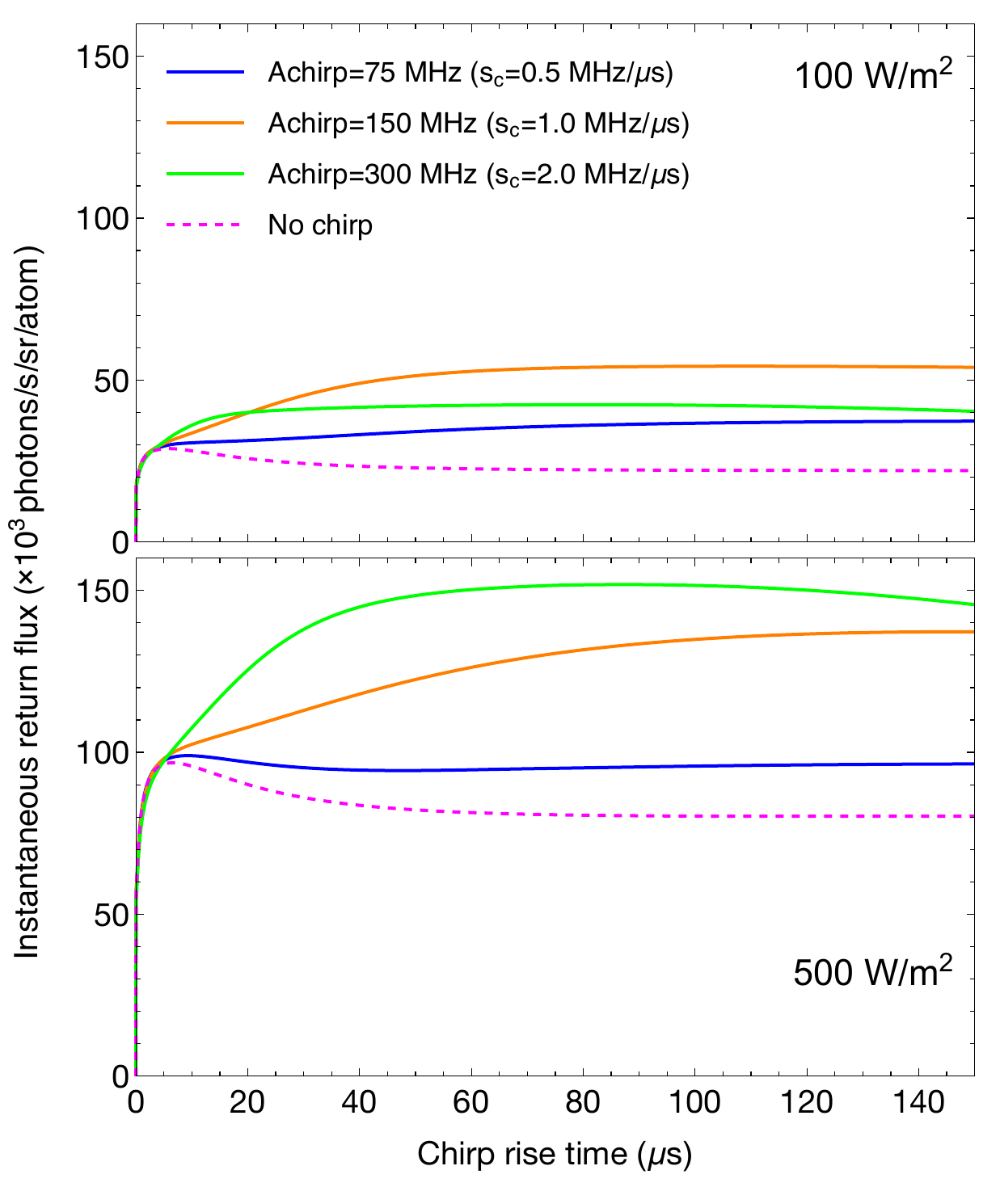} \\
    \caption{Time evolution of the return flux for excitation with irradiance of 100~W/m$^2$ (top) and 500~W/m$^2$ (bottom), during a chirping period $\tau_\text{chirp}=150$~$\mu$s and for several chirping amplitudes including the case of fixed frequency excitation (no chirp). The chirping rate or slope ($s_c$) is indicated for each chirping amplitude.}  \label{fig:return_evolution}
\end{center}
\end{figure}

\subsection{Evolution of atomic population during chirping}

In order to boost pumping of sodium, the optical frequency of light can be periodically swept (chirped). The chirping is characterized by the amplitude or frequency range $A_\text{chirp}$ of the sweep, and the rise time or period of the chirp $\tau_\text{chirp}$. We define the chirp rate or slope by $s_c\vcentcolon = A_\text{chirp}/\tau_\text{chirp}$. It is the purpose of this section to understand how the atomic population, and ultimately the emission of photons, depends on the chirp parameters. 

Figure \ref{fig:velocity_sequence} shows a sequence of the evolution of the velocity distribution for chirp parameters $\tau_\text{chirp}=150$~$\mu$s and $A_\text{chirp}=(75\,,100\,,300)$~MHz. The ground-state population (blue), excited-state population (orange), and laser frequency detuning (red vertical line) are depicted for every given time. As a reference, the evolution with fixed frequency pumping (no chirping) is shown at the bottom of the figure.  
At $t=0$, the optical frequency is tuned to $\omega=\omega_0-A_\text{chirp}/2$. Here, a group of atoms in a blueshifted velocity class with respect to $\omega_0$ is resonant with the incoming photons and a bump starts to build up in the following velocity class due to recoil. Initial conditions are the same in all cases, except for the different absorption cross section for each starting laser detuning. Due to atoms being pumped to the $\ket{F'=2, m'=+2}$ excited state (where $m$ is the magnetic quantum number) by the laser repumping sideband ($q=0.1$ in all cases), there is a small contribution to fluorescence in a group with a detuning of $+60$~MHz with respect to the laser frequency. 

The laser frequency $\omega$ is increased, tracking the atomic population. At $t=\tau_\text{chirp}/2=75$~$\mu$s, the laser frequency is exactly $\omega_0$. At this point, the absorption cross section is the highest and more atoms are resonant with the laser, increasing saturation and the number of recoiled atoms. The large peak in the ground-state distribution for $A_\text{chirp}=75$~MHz indicates a suboptimal chirping rate with respect to $A_\text{chirp}=150$~MHz and $300$~MHz, because a large fraction of atoms are pushed to higher non-resonant velocity groups. For this case, the chirp rate is too low. With $A_\text{chirp}=150$~MHz, the peak and the dip observed in the ground-state distribution are smaller. The contribution to fluorescence, i.e. the excited-state distribution, is larger when the chirped laser can more efficiently follow the shifting atomic population. The center of the hole appears here behind the current laser frequency detuning, accounting for relaxation mechanisms (collisions) slower than the chirp rate. For $A_\text{chirp}=300$~MHz, the hole and the peak are the smallest of all cases, although fluorescence is still larger than in the no chirping scenario. The chirping rate is so high that saturation and therefore optical pumping on each velocity class does not develop efficiently, although the atoms that recoil can be re-excited by the laser. The vestiges of the initial hole burning at $-150$~MHz are barely visible as a consequence of the relaxation mechanisms taking place during the rapid chirp.

At $t=\tau_\text{chirp}=150$~$\mu$s, only the ground-state distribution with $A_\text{chirp}=75$~MHz shows a slight increase on its peak, while other cases seem to reach a steady state. Raising the laser frequency further would decrease the efficiency of the process as the absorption cross section becomes smaller. Indeed, this is the reason why the range of chirp frequencies is symmetric with respect to the central velocity class.

The evolution of the instantaneous return flux (fluorescence opposite to the direction of laser propagation) along $\tau_\text{chirp}$ for $I=(100\,,500)$~W/m$^2$ is shown in Fig.~\ref{fig:return_evolution} for the same chirping (and non-chirping) parameters shown in Fig.~\ref{fig:velocity_sequence}. For $I=100$~W/m$^2$, the largest return flux is obtained with $A_\text{chirp}=150$~MHz ($s_c=1$~MHz/$\mu$s), more than twofold higher than the steady state flux in the case of no chirping. As observed in the velocity-distribution sequence for the case of $A_\text{chirp}=75$~MHz, even at $t=150$~$\mu$s the return flux is slightly increasing towards some steady state. 
Increasing the irradiance on the sodium layer always raises the return flux, although the efficiency of absorption becomes smaller. For $I=500$~W/m$^2$ the largest return flux among the simulated cases is obtained with $A_\text{chirp}=300$~MHz ($s_c=2$~MHz/$\mu$s), although the relative enhancement with respect to the return flux with fixed frequency is smaller than what can be obtained at $I=100$~W/m$^2$. The largest return occurs with a rapid chirp because saturation occurs faster in the presence of higher irradiance and therefore less time needs to be spent to saturate the velocity class along the frequency sweep. 

These two examples indicate that the optimal chirp parameters depend on the irradiance level, and the observation that steady state is reached for different values of the chirp period introduces the chirp rise time as an additional degree of freedom. 

\begin{figure*}[h]
\begin{center}  
     \includegraphics[width=0.23\linewidth]{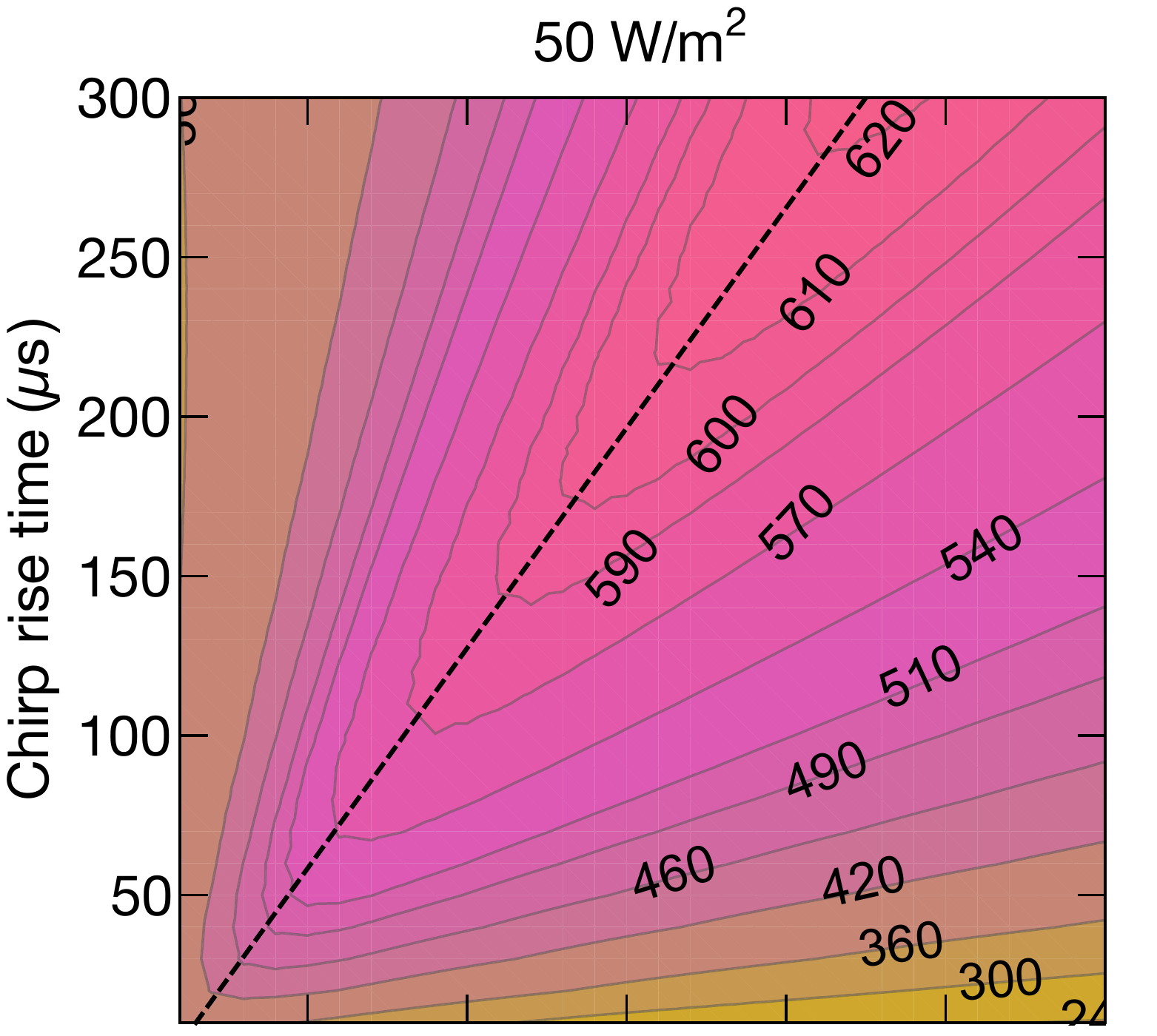} \hspace{-0.8cm}
       \includegraphics[width=0.23\linewidth]{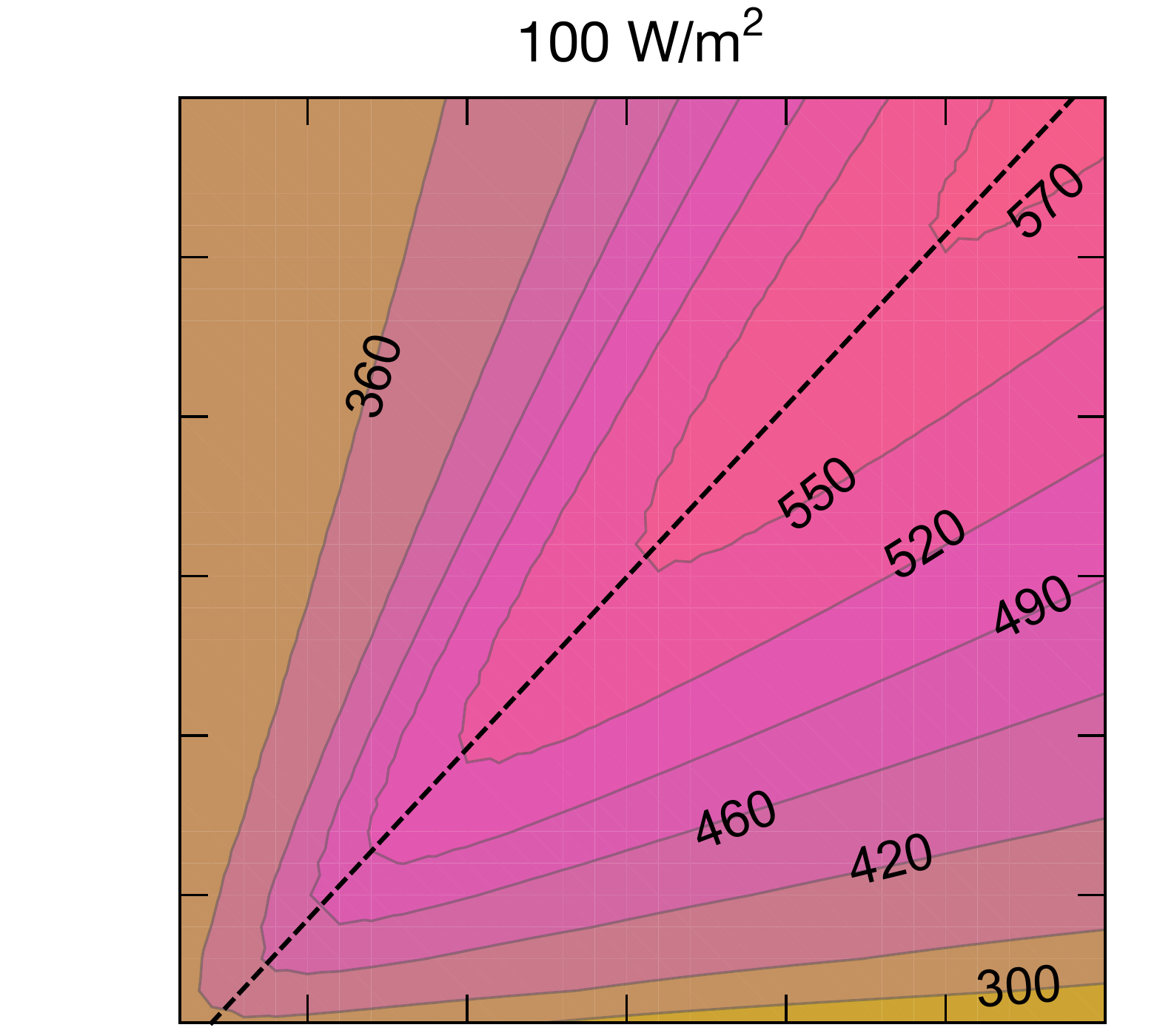} \hspace{-0.8cm}
       \includegraphics[width=0.23\linewidth]{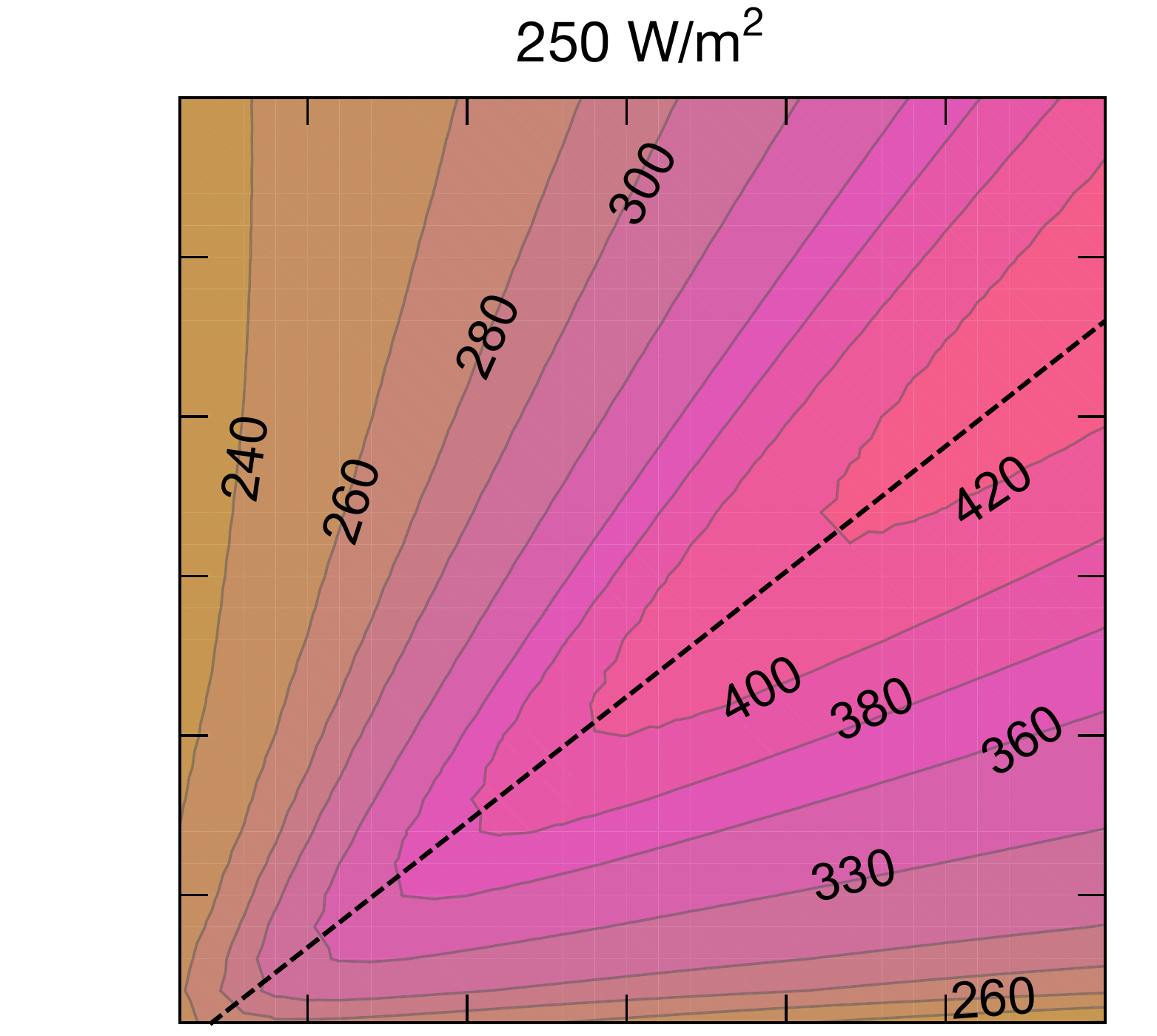} \hspace{-0.8cm}
       \includegraphics[width=0.23\linewidth]{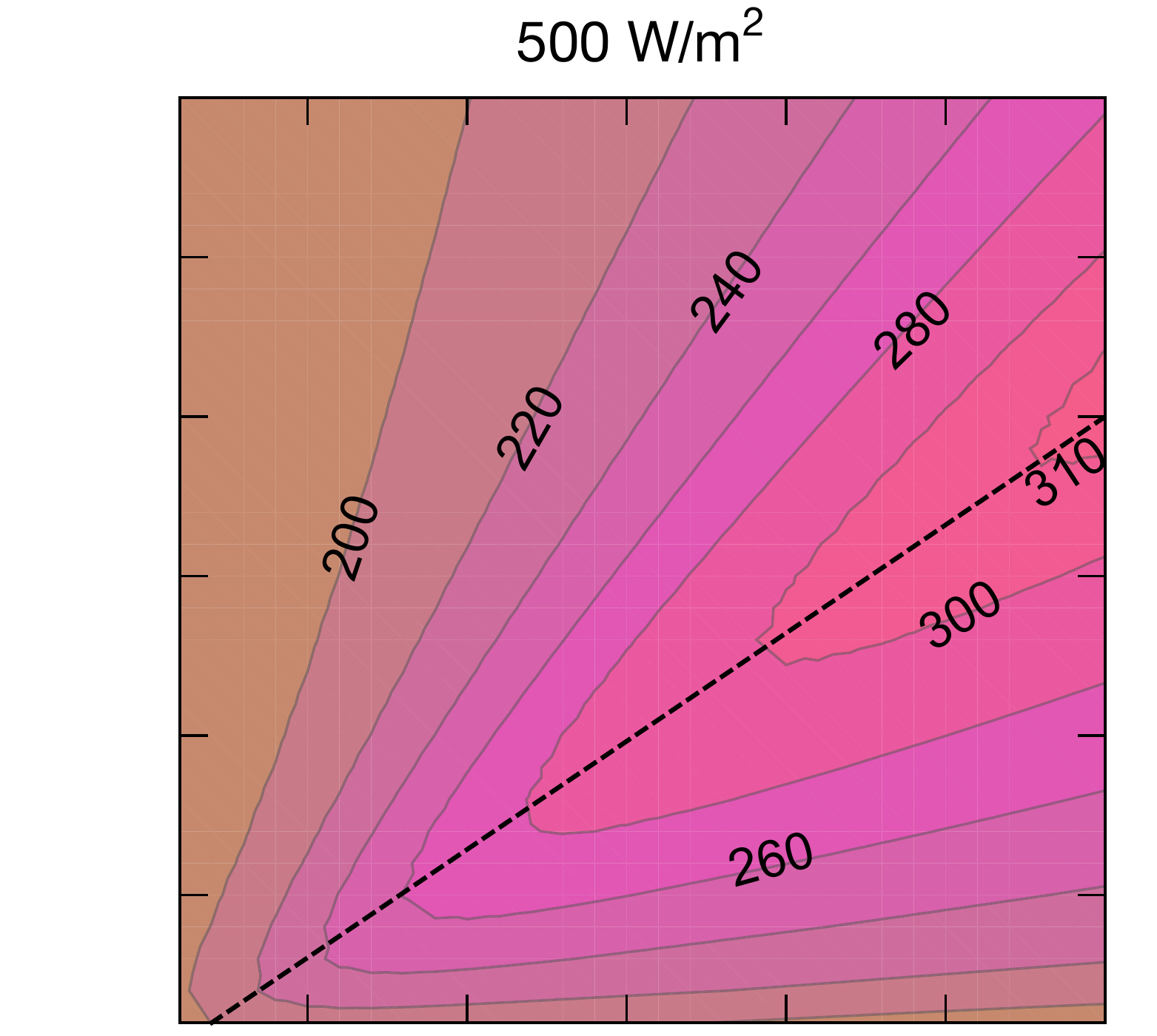}\\
       \hspace{-0.1cm}
     \includegraphics[width=0.23\linewidth]{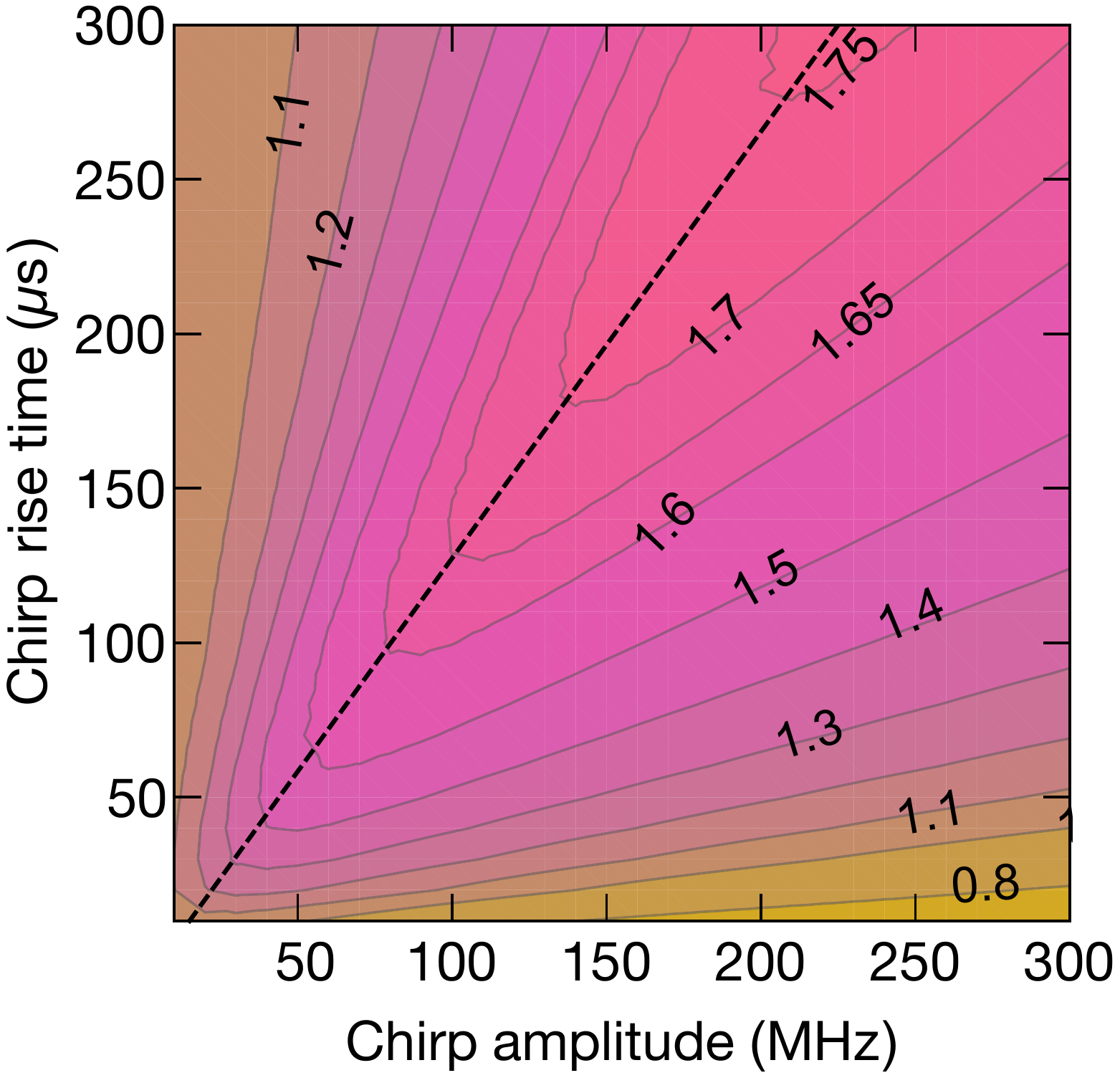} \hspace{-0.8cm}
       \includegraphics[width=0.23\linewidth]{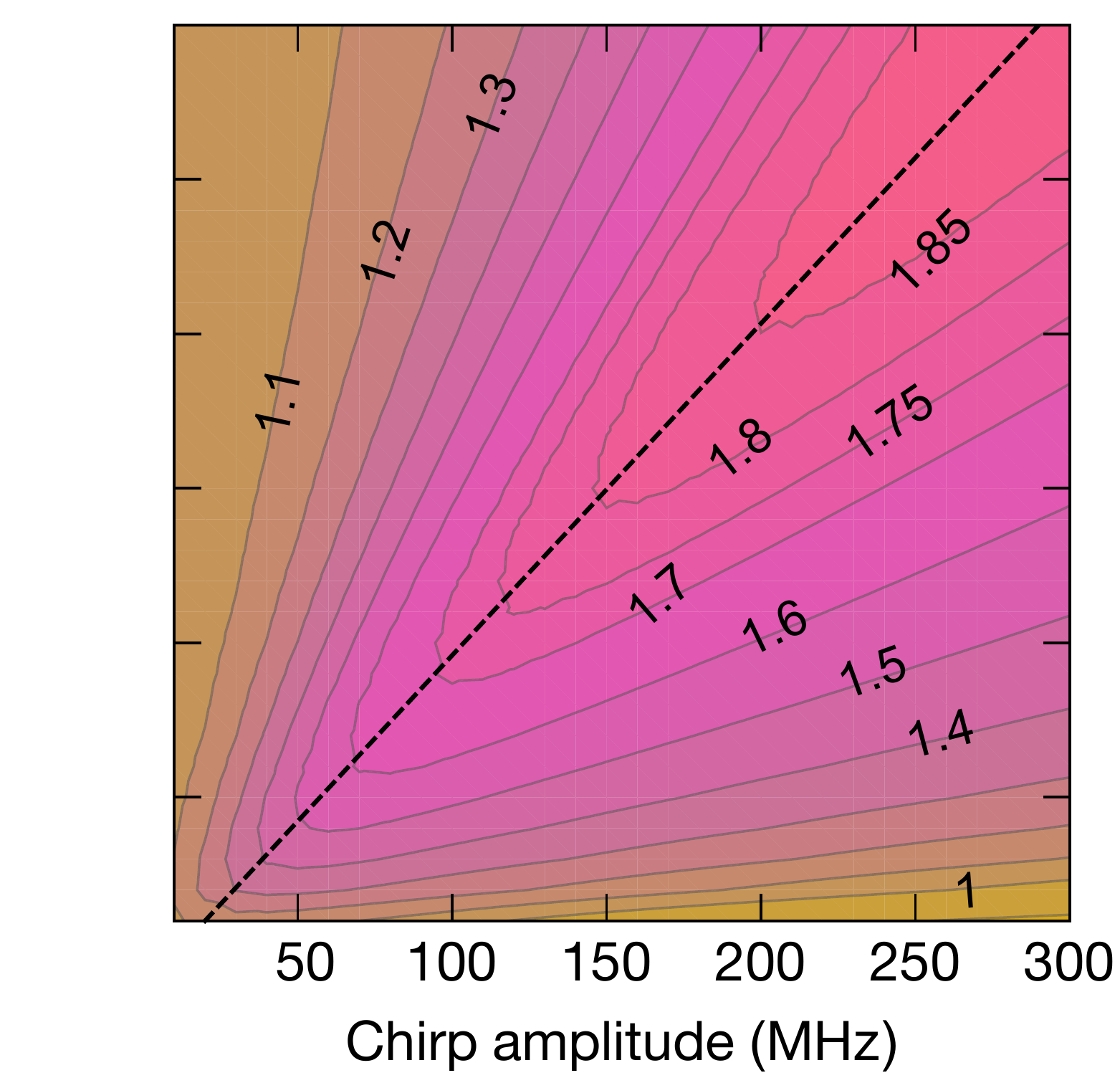} \hspace{-0.8cm}
       \includegraphics[width=0.23\linewidth]{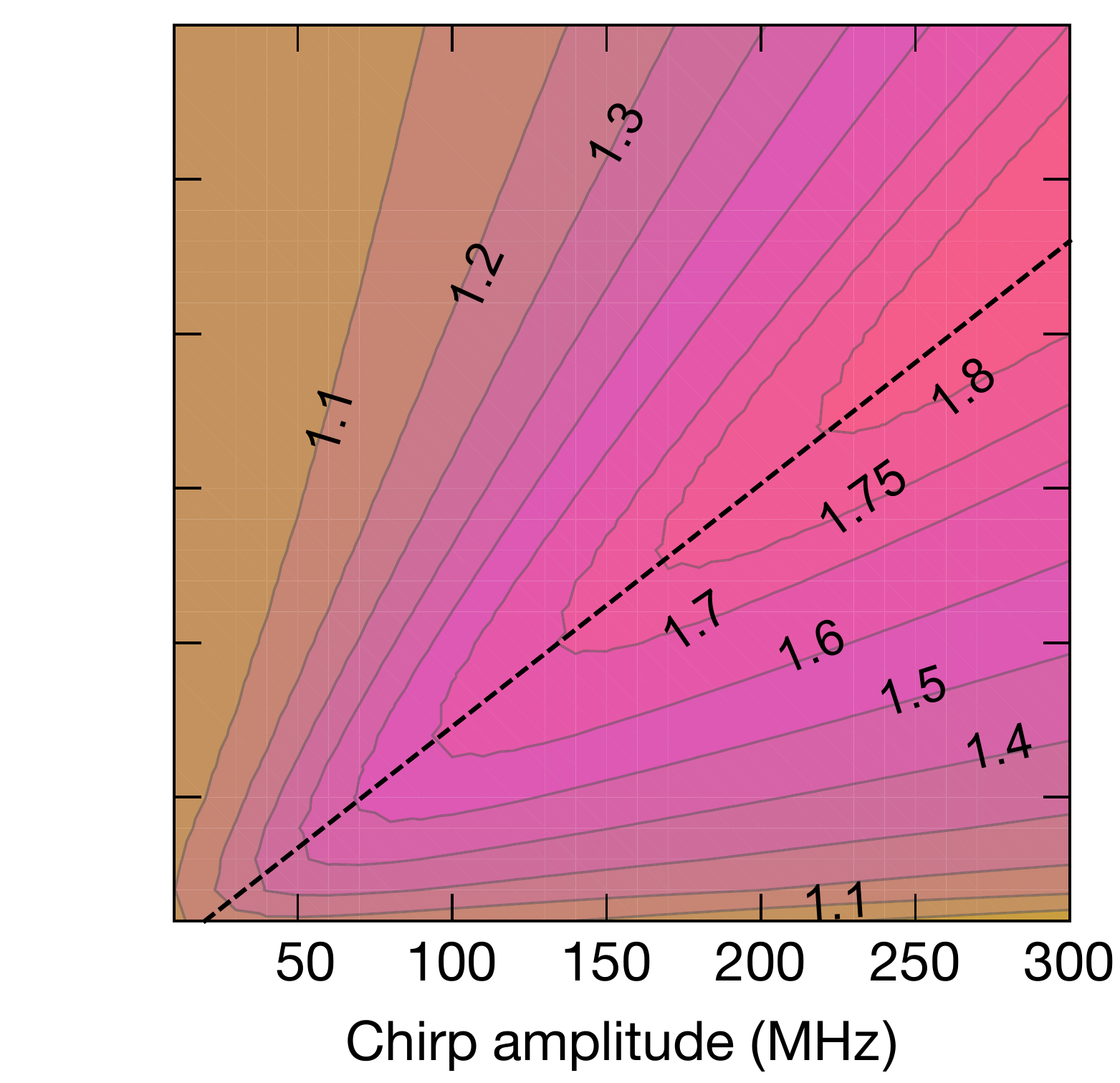} \hspace{-0.8cm}
       \includegraphics[width=0.23\linewidth]{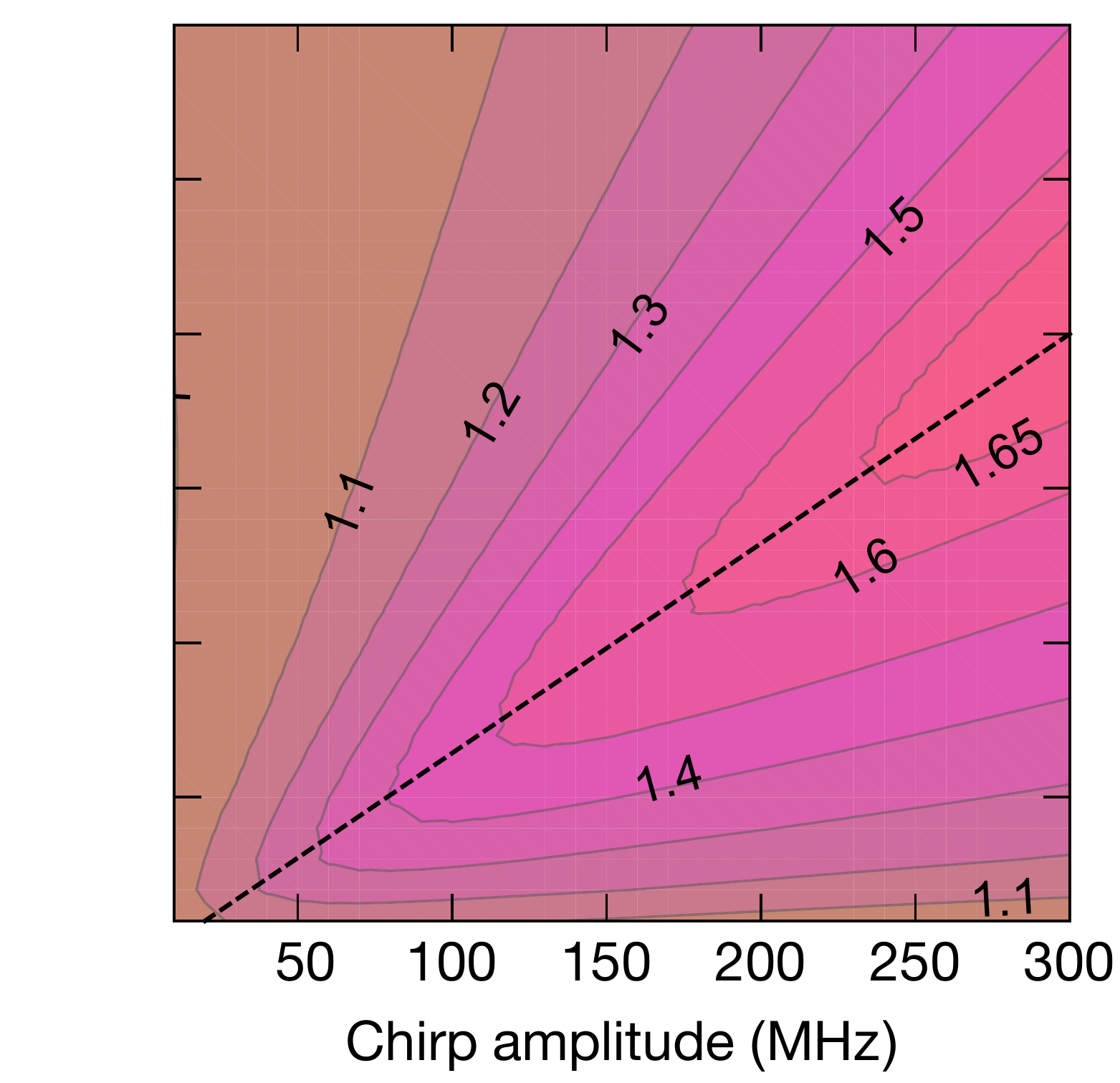}
    \caption{Specific return (top) and enhancement factor $\kappa$ (bottom) as a function of the chirping parameters $A_\text{chirp}$ and $\tau_\text{chirp}$ for a set of irradiances (indicated above the plots) and assuming standard parameters, but $\theta_B=0$\degree. A linear function with the slope of maximum return (dashed line) is shown on every plot.}  \label{fig:chirp_AcTc}
\end{center}
\end{figure*}

\subsection{Chirp rate}


With an optimal chirp rate, a balance between high optical pumping and quick tracking of the shifting atomic populations must be reached. Figure~\ref{fig:chirp_AcTc} shows the simulated specific return $\psi$ in units of (photons/s/sr/atom)/(W/m$^2$) and the enhancement factor $\kappa$ for several irradiance values, as a function of the frequency range and rise time for $A_\text{chirp}=10$--$300$~MHz and $\tau_\text{chirp}=10$--$300$~$\mu$s, respectively. The enhancement factor or gain $\kappa$ is here introduced and defined as the ratio between the specific return with chirped light $\psi_\text{chirping}$ and the specific return without chirp (fixed frequency at zero detuning) $\psi_\text{no-chirping}$, such that
\begin{equation}
\kappa =\dfrac{\psi_\text{chirping}}{\psi_\text{no-chirping}}.
\end{equation}  

For each irradiance value shown in Fig.~\ref{fig:chirp_AcTc} we calculate a linear slope function over which the optimal performance (return flux and enhancement) can be obtained. The maximum performance is, however, only achieved at a particular point over this slope function. The optimal set of parameters $A_\text{chirp}$ and $\tau_\text{chirp}$ is typically constrained to experimental factors. Generally speaking, a long $\tau_\text{chirp}$ and a small $A_\text{chirp}$ would be easier to implement in a real laser system.   

The maximum specific return is obtained at $I=50$~W/m$^2$ at the chirp slope $s_c=0.7$~MHz/$\mu$s, while the maximum enhancement $\kappa=1.85$ is found at $I=100$~W/m$^2$ with $s_c=1.0$~MHz/$\mu$s. With higher light intensities the gain of chirping is slowly reduced, which is attributable to power broadening effects as shown in Fig.~\ref{fig:velocity_Maxwell}.
    
\begin{figure}[h]
\begin{center}  
   \includegraphics[width=0.8\linewidth]{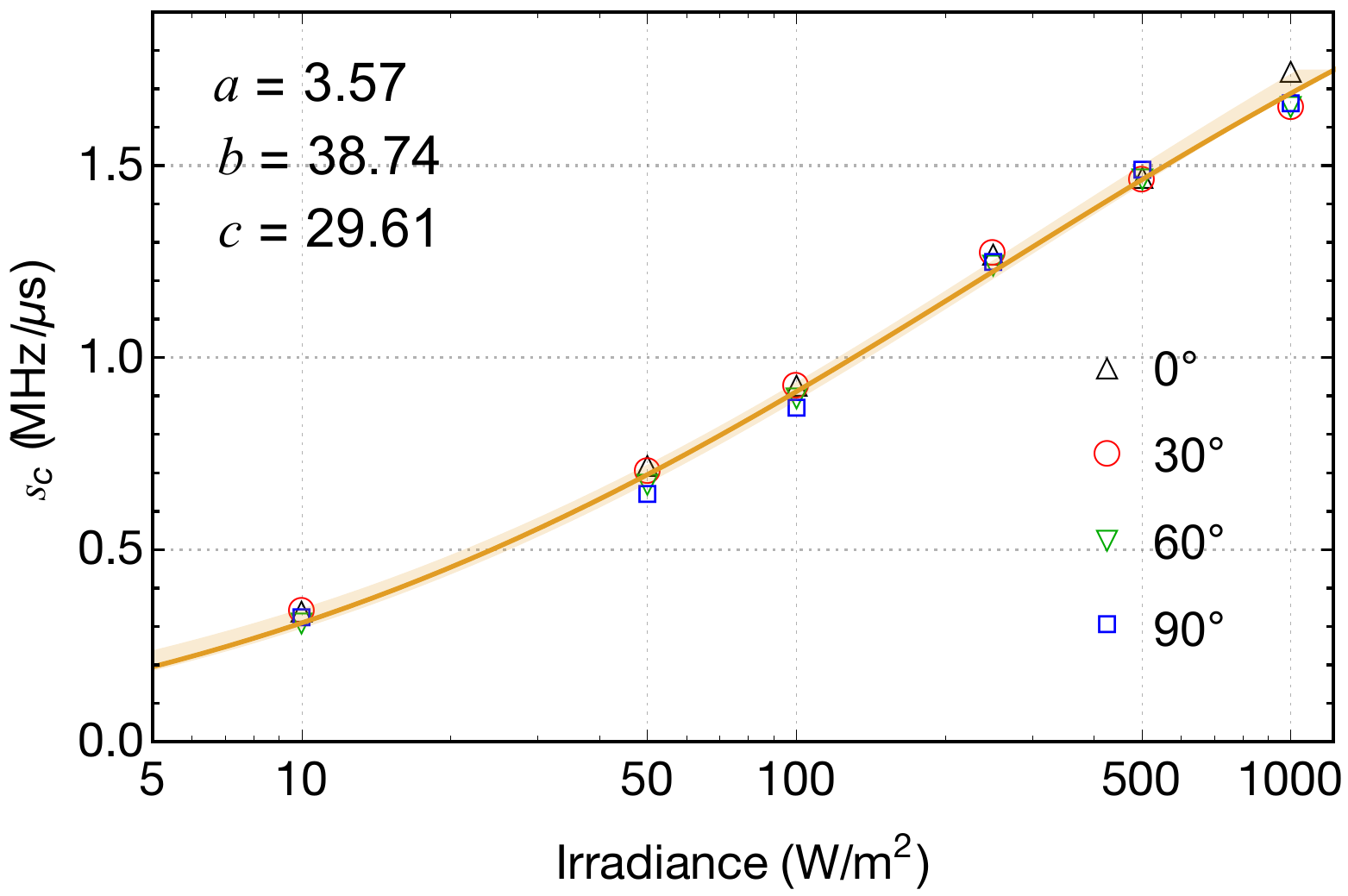} \\
    \caption{Calculated chirping slope ($s_c$) of maximum return as a function of irradiance for $\theta_B=(0\,,30\,,60\,,90)$\degree\ with the fit function defined in Eq.~\ref{eq:fit_sc} (brown curve) and the corresponding averaged fit parameters $a$, $b$ and $c$.}  \label{fig:sc_maximum_return}
\end{center}
\end{figure}

The optimal chirping slope at which maximum return is obtained can be characterized with the following empirical expression:
\begin{equation}
s_c(I) = a-\exp{\left( \frac{b}{c+\sqrt{I}} \right)}, \label{eq:fit_sc}
\end{equation}
where $a$, $b$, and $c$ are fit parameters, $I$ is the irradiance in W/m$^2$ and $s_c$ is expressed in MHz/$\mu$s.
The optimal chirping slopes found in Fig.~\ref{fig:chirp_AcTc} are calculated for $\theta_B=0$\degree, although we also simulated for $\theta_B=(30\,,60\,, 90)$\degree\ and up to $I=1000$~W/m$^2$. All chirping slopes are shown in Fig.~\ref{fig:sc_maximum_return} where we calculated a slope fit function for each curve. We find that there is no significant difference in the optimal chirping rate as a function of the magnetic polar angle $\theta_B$, as it is shown in Fig.~\ref{fig:sc_maximum_return} in a yellow band. The mean value of the fit parameters are $a=3.57$, $b=38.74$, and $c=29.61$, for $s_c$ and $I$ measured in MHz/$\mu$s and $W/$m$^2$, respectively.  



\subsection{Repumping}

\begin{figure}[h]
\begin{center}  
   \includegraphics[width=0.99\linewidth]{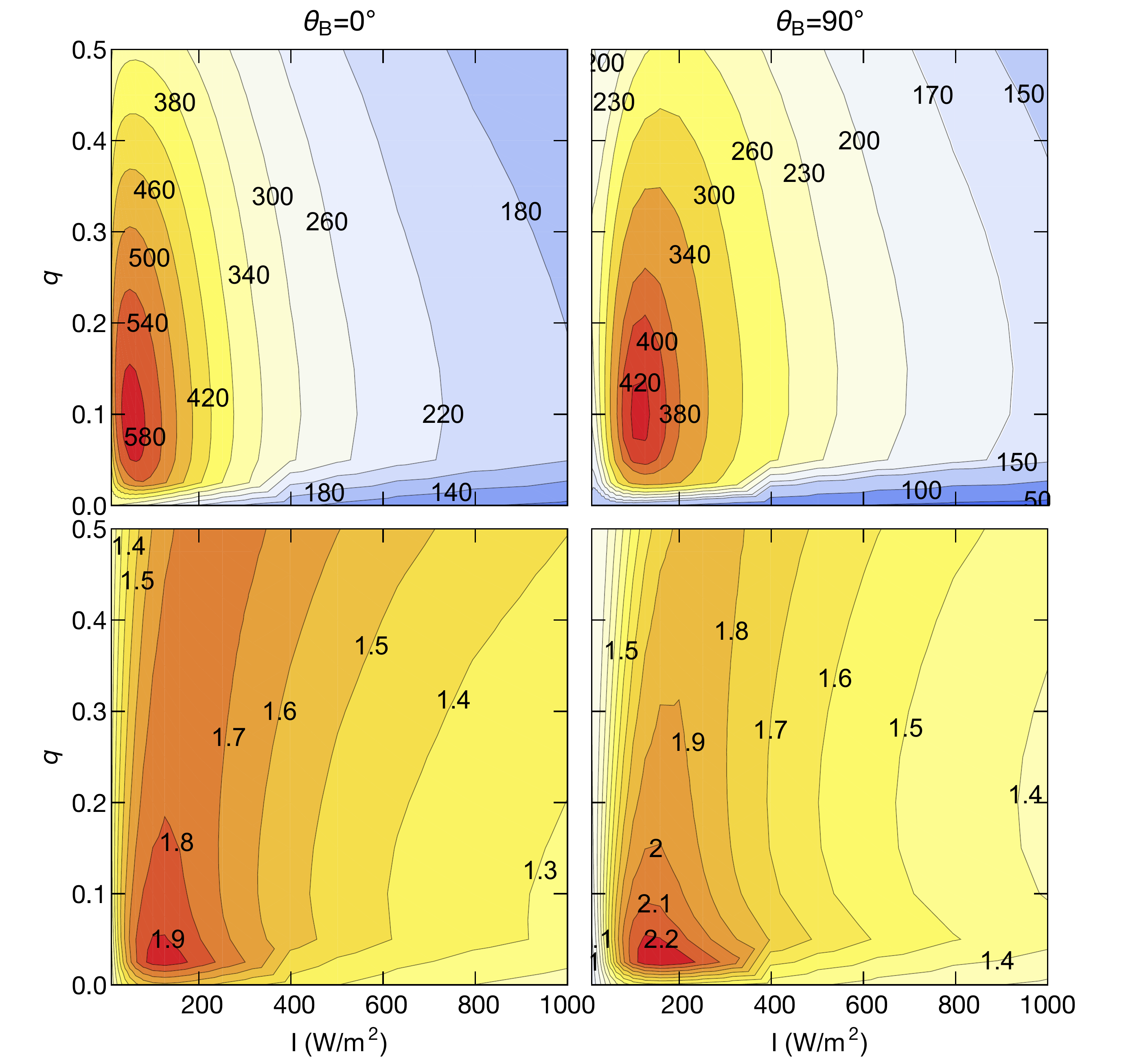}
    \caption{Specific return flux $\psi_\text{chirp}$ (top) and enhancement factor $\kappa$ (bottom) as a function of the irradiance and repumping fraction $q$. Magnetic field is 0.36~G and chirping parameters are $A_\text{chirp}=200$~MHz and $\tau_\text{chirp}=200$~$\mu$s.}  \label{fig:Repumping}
\end{center}
\end{figure}

It has been found that the optimal repumping fraction $q$, defined in Eq.~\ref{eq:repumping}, lies between 0.1 to 0.15 for 20 W-class guidestar lasers and the question is whether the same applies for the case of laser chirping. Figure~\ref{fig:Repumping} shows the specific return flux and the enhancement for a range of irradiance and repumping fractions. In this case, the magnetic field is taken into account in the model ($B=0.36$~G) and we plot for the two cases of the angle between the laser and the magnetic field of $\theta_B=0$\degree\ and $\theta_B=90$\degree. The maximum return flux efficiency with chirped light is obtained for $q$ near 0.1 (10\%) for both cases and near an irradiance of 100~W/m$^2$. This irradiance is expected as it corresponds to the optimal point for a chirping rate of 1~MHz/$\mu$s, according to Eq.~\ref{eq:fit_sc}. The maximum enhancement can be found for repumping fractions around $q=0.05$ (5\%). Even though this does not lead to the maximum return, it shows that the maximum efficiency of the chirping mechanism, compared to the traditional fixed frequency pumping, can be obtained with a rather small $q$.

\begin{figure*}[t]
\begin{center}  
   \includegraphics[width=0.9\linewidth]{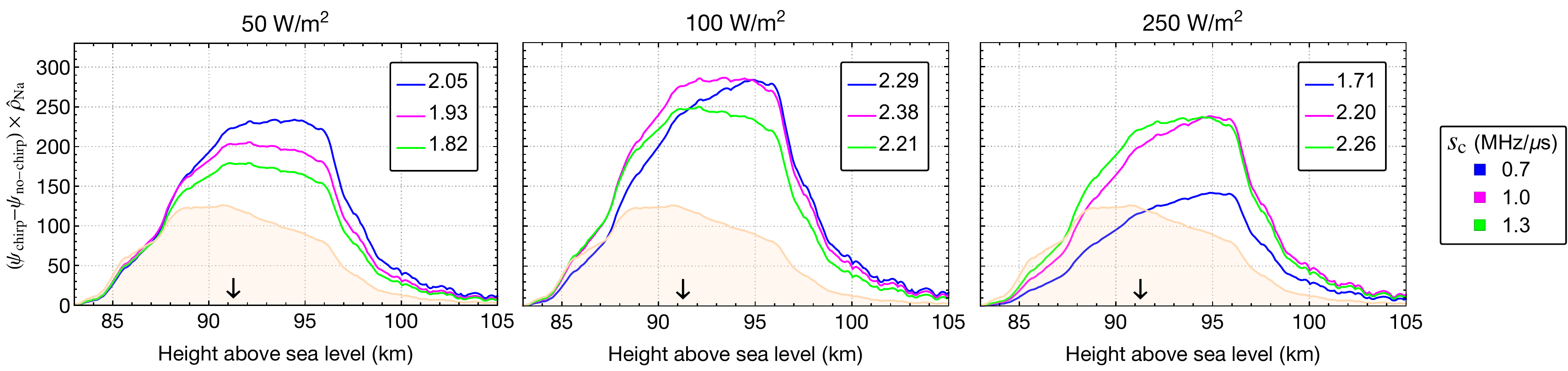} 
    \caption{Return flux contributions due to laser chirping along the vertical profile in the mesosphere. The enhancement factor for each chirp rate are shown in the legend of each plot. The sodium profile (light orange) is presented for reference on an arbitrary scale along with a vertical arrow to indicate the position of the sodium centroid. Magnetic field is 0.36~G and $\theta_B=45$\degree.}  \label{fig:profile_return}
\end{center}
\end{figure*}

\subsection{Return flux}

So far all calculations have been performed for fixed collision rates. To obtain a closer answer to the actual return flux with the chirping technique, we include in the calculations vertical profiles of atomic densities and temperature, leading to the collision rates shown in Fig.~\ref{fig:relaxation}.(c). To quantify the return flux contribution of the chirping method with respect to a fixed-frequency LGS, we calculate the specific return along the vertical profile between 83~km and 105~km with the corresponding collision rates. Then, we take the product of a normalized reference sodium profile with the difference between the chirped and non-chirped specific-return profiles. In this way, we can estimate where in the vertical profile the additional contribution to the overall fluorescence due to chirping is maximum. Results of applying this method to mesospheric irradiance of 50~W/m$^2$, 100~W/m$^2$ and 250~W/m$^2$ are shown in Fig.~\ref{fig:profile_return} for several chirp rates. The total return flux can be estimated by integrating the specific-return profiles weighted with the sodium profile. The enhancement factor is then calculated as the ratio of the integrated profiles. For each curve in Fig.~\ref{fig:profile_return}, the enhancement factor is shown as a legend on each panel. 

Several observations can be made from these simulations. The chirping slope at which the maximum enhancement occurs depends on the irradiance in the same way as seen before, that is, for higher irradiance a higher chirp rate is desired. There is a trend to obtain more return at higher altitudes with increasing irradiance. Although all curves at $I=250$~W/m$^2$ suddenly drop at approximately 96~km altitude due to the drastic reduction in sodium density, the increasing return flux contribution from laser chirping is explained by the fact that at high altitude collision are fewer and more atoms can be ``snow-ploughed'' before they dissipate in the velocity space. Indeed, it is likely that a rise in the sodium centroid maximizes the benefits of laser chirping. 

Finally, we perform a comprehensive simulation of the expected return flux of an LGS that would be measured on the ground, taking the location of La Palma as a representative site of an astronomical observatory, motivated by ongoing experimental tests on laser chirping at the Observatorio del Roque de los Muchachos (ORM) in La Palma. We assume a near-collimated laser beam of 30~cm launched diameter which is propagated through the atmosphere and sodium layer. We discretize the vertical profiles in 23 layers and calculate the return flux contribution for each. The Gaussian beam cross section is divided in contours of six irradiance levels and the calculation of the specific return is performed on each of them. After integration over the beam profile, we obtain the product of the beam integration $B_i$ with a discretized sodium column density $C_{\text{Na},i}$ pertaining to the reference sodium profile, which results in the photon flux contribution on each layer $i$ in the vertical direction. The total photon flux $\Phi$ in units of (photons/s/m$^2$) on the detector can be estimated as
\begin{equation}
\Phi = \sum_{i=1}^N \dfrac{X \eta^X C_{\text{Na},i} B_i}{L^2},
\end{equation} 
where $\eta$ is the one-way vertical atmospheric transmission, $X=\sec(\xi)$ is the airmass and $\xi$ is the zenith angle, $L$ is the distance between the telescope and the sodium centroid, and $N$ is the number of layers.

\begin{figure}[t]
\begin{center}  
   \includegraphics[width=0.98\linewidth]{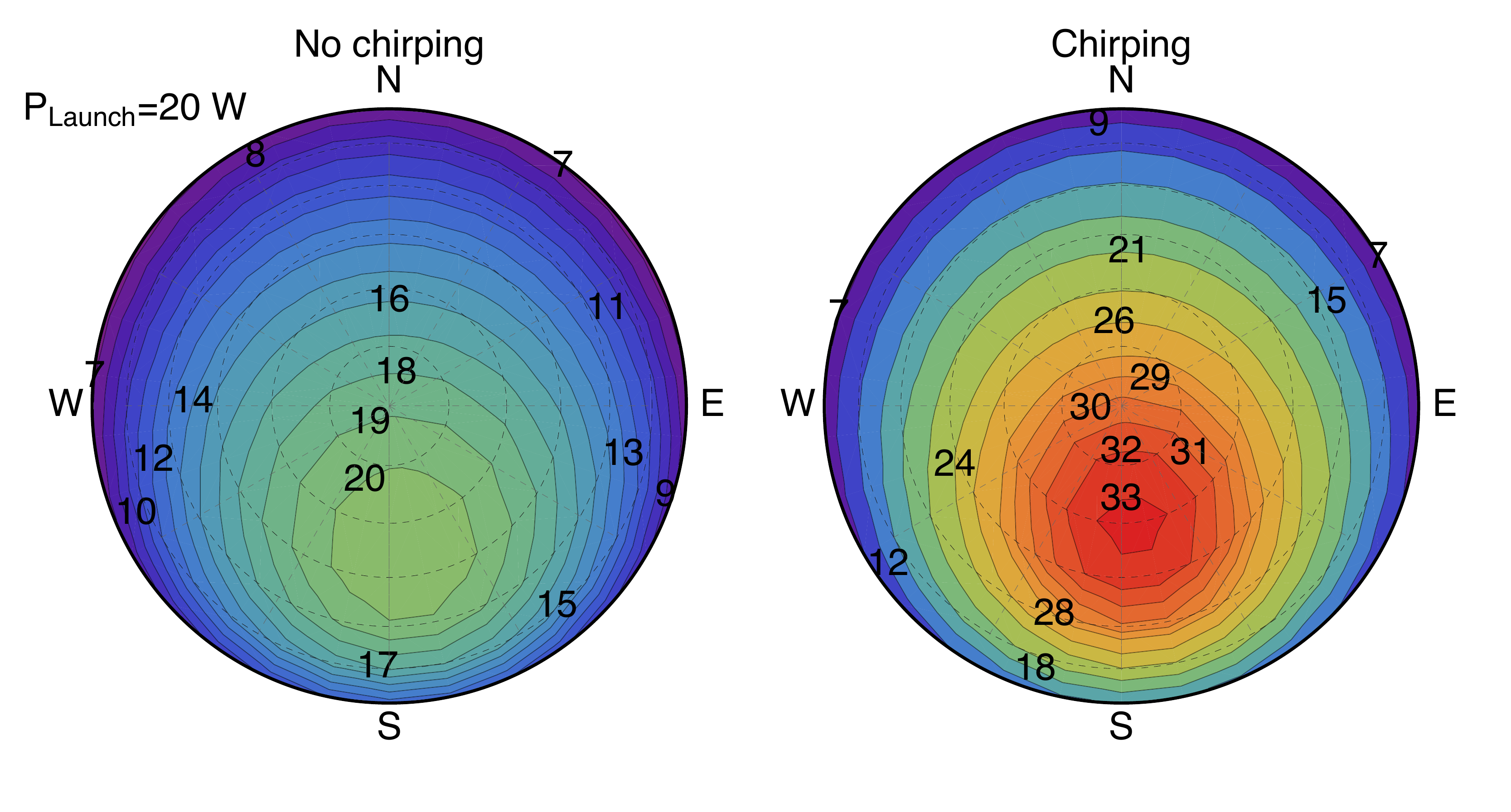} \\
   \includegraphics[width=0.98\linewidth]{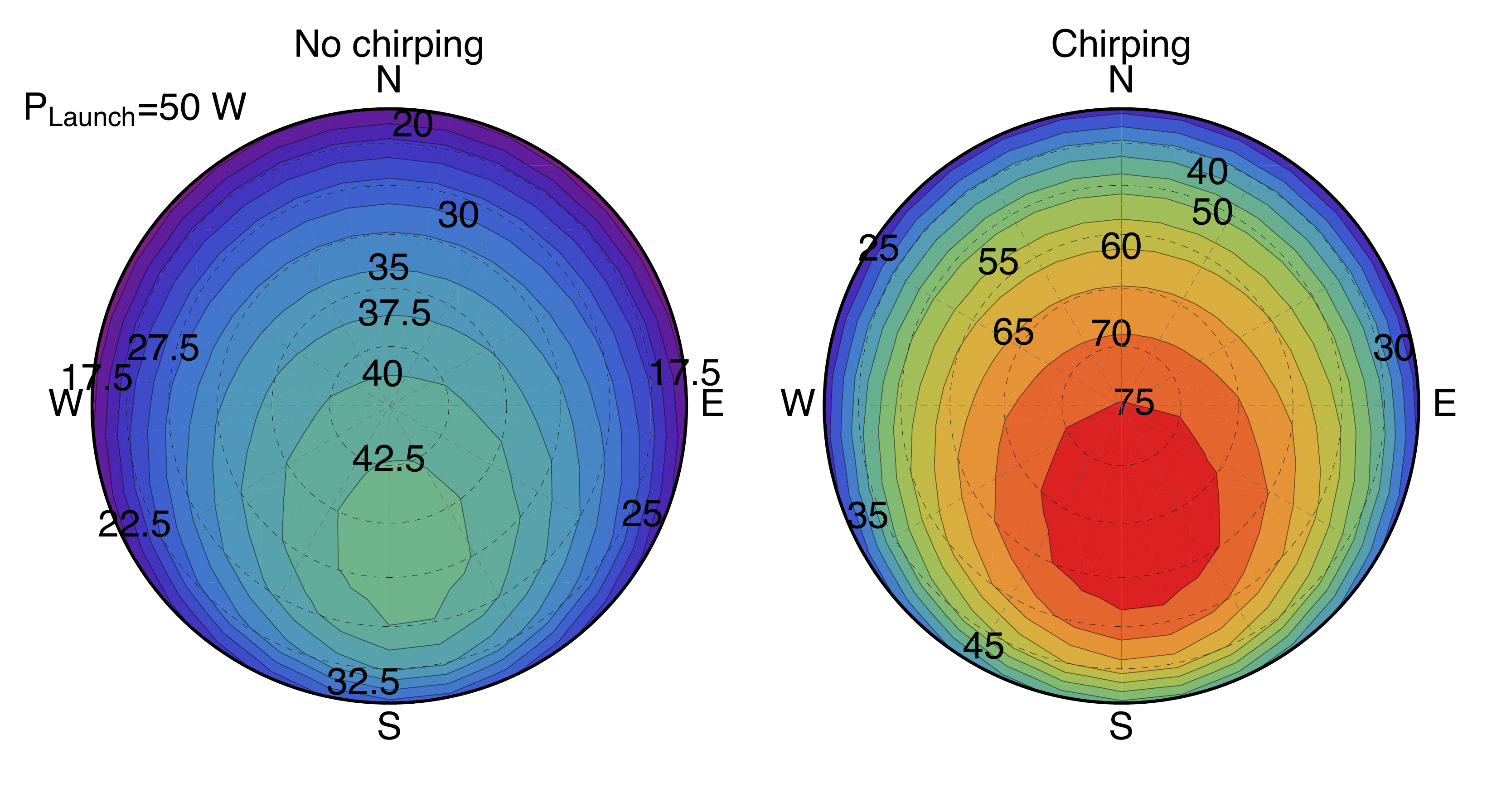}  
    \caption{Sky plots comparing the return flux on the ground $\Phi$~($10^6$~photons/s/m$^2$) obtained with CW guidestar lasers with and without chirping above La Palma, for launched laser power of 20~W and 50~W and chirp parameters $A_\text{chirp}=250$~MHz and $\tau_\text{chirp}=300$~$\mu$s ($s_c=0.83$~MHz/$\mu$s).}  \label{fig:SkyPlot_LaPalma}
\end{center}
\end{figure}

Figure~\ref{fig:SkyPlot_LaPalma} shows a series of sky plots with contour lines indicating the photon return flux in units of $10^6$~photons/s/m$^2$ received if the laser is pointed in the corresponding direction in the sky for a given laser power. The sky plots cover the full azimuth range and a zenith angle up to 60\degree\. We perform a simulation with and without chirping for a launched laser power of 20~W and also for 50~W single-frequency guidestar lasers. Results obtained for fixed frequency excitation are consistent with those from simulations and experiments in Tenerife \citep{Holz:2016}, which due to its close proximity and almost same elevation as La Palma, has the same experimental parameters. For both laser formats, the maximum return flux is obtained when pointing in a direction where the laser is parallel to the geomagnetic field lines ($\theta_B=0\degree$). In other directions the return flux is reduced due to Larmor precession, although chirping can still raise the increasing return flux with respect to the fixed frequency format. The predicted return flux enhancement with laser chirping is substantial, particularly with a 50~W class laser. 

\begin{figure}[t]
\begin{center}  
   \includegraphics[width=0.48\linewidth]{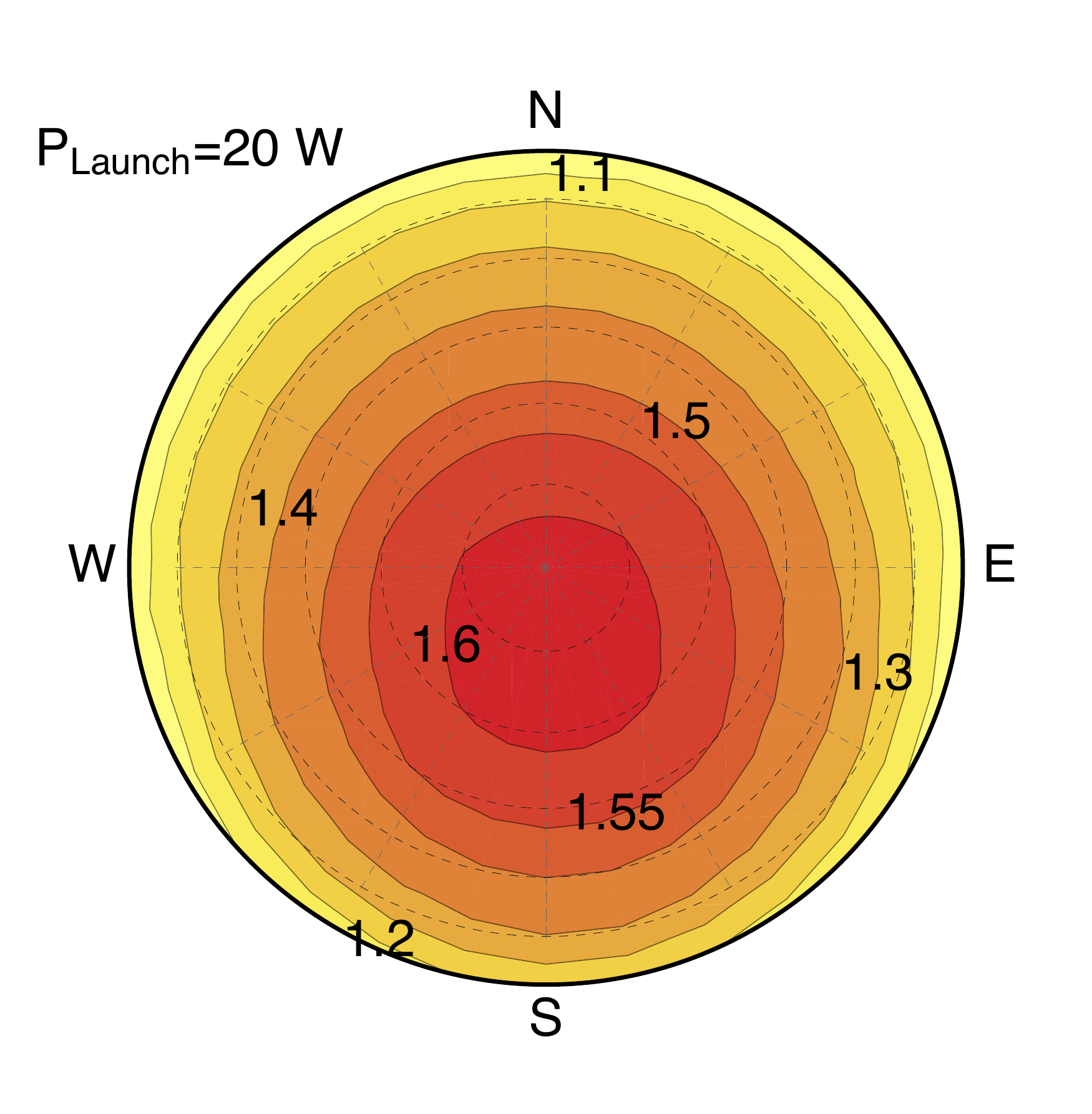}
   \includegraphics[width=0.48\linewidth]{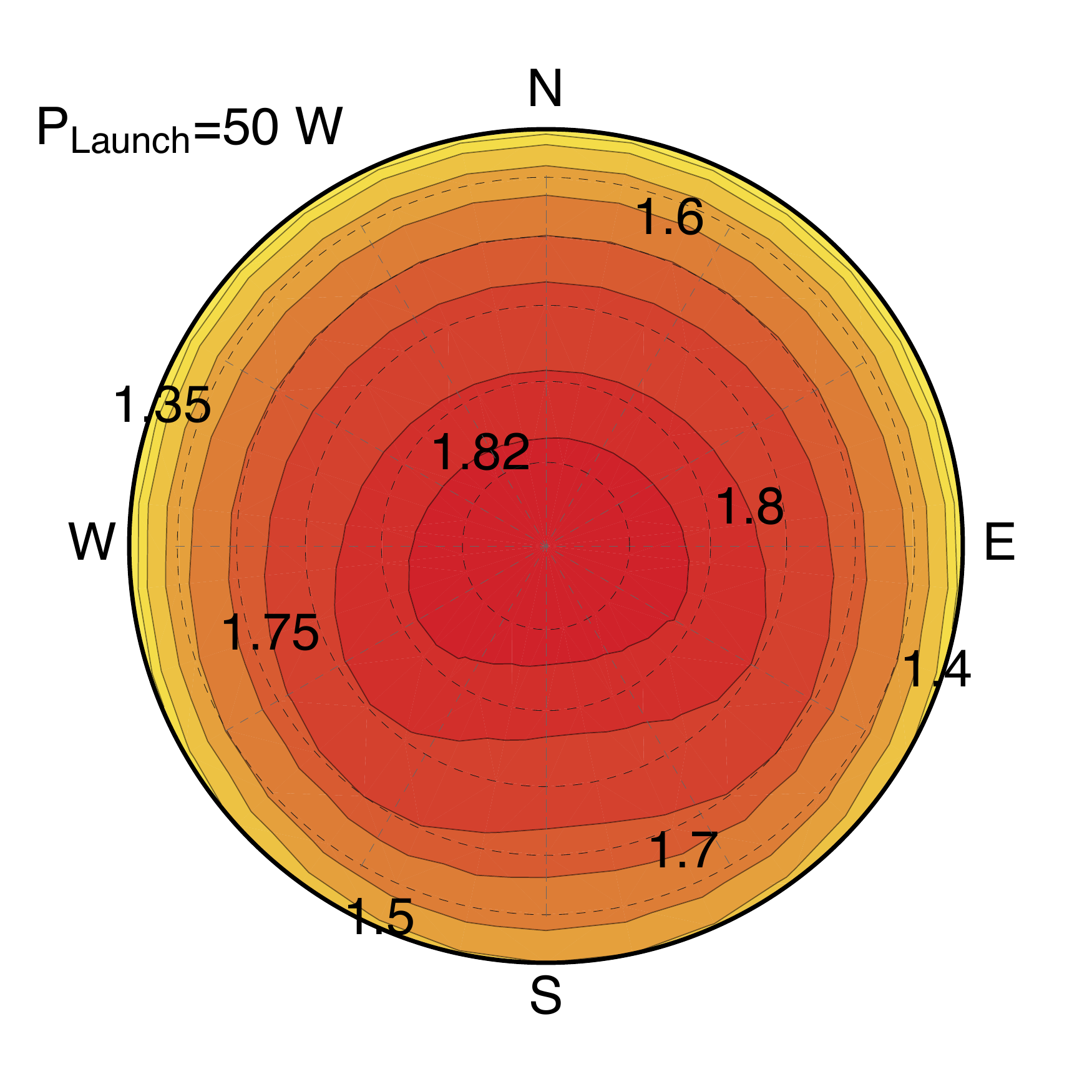}  
    \caption{The enhancement factor $\kappa$ for the two examples shown in Fig.~\ref{fig:SkyPlot_LaPalma}. An almost twofold enhancement in return flux could be achieved with future 50~W class CW guidestar lasers.}  \label{fig:SkyPlot_Gain_LaPalma}
\end{center}
\end{figure}

Sky plots with the enhancement factor $\kappa$ obtained from these simulations are shown in Fig.~\ref{fig:SkyPlot_Gain_LaPalma}, where an enhancement up to 82\% (factor of 1.82) can be obtained with a 50~W laser and even more than 30\% at high zenith angles. It is interesting to note that the direction of maximum enhancement tends to deviate to the north (towards larger $\theta_B$) with increasing laser power. 

\begin{figure}[t]
\begin{center}  
   \includegraphics[width=0.8\linewidth]{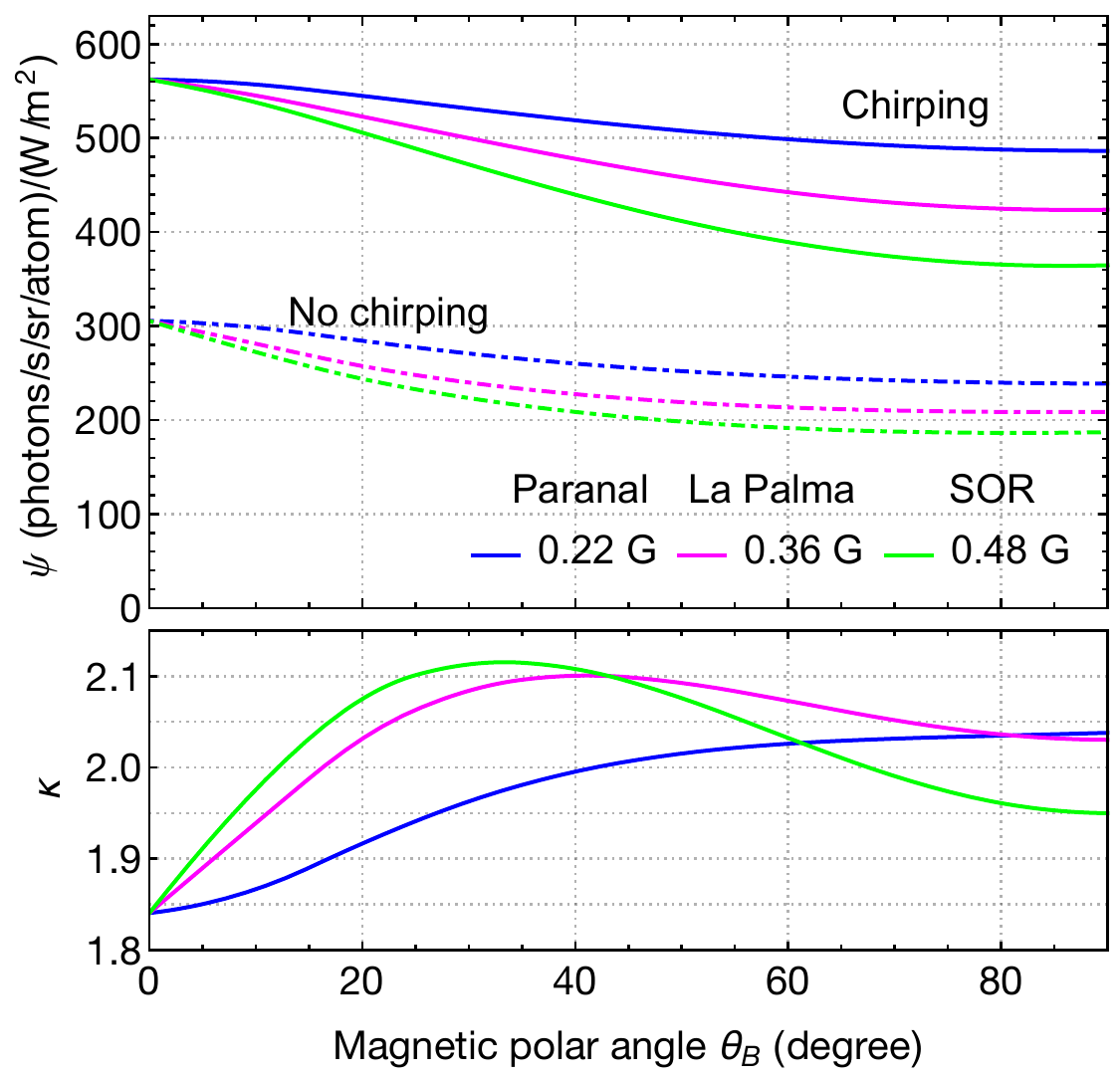} \\
    \caption{Specific return $\psi$ and enhancement factor $\kappa$ as a function of the magnetic polar angle $\theta_B$ for different strenght of magnetic field (Paranal: 0.22~G, La Palma: 0.36~G, SOR: 0.48~G). $I=100$~W/m$^2$, $s_c=1.0$~MHz/$\mu$s with $A_\text{chirp}=200$~MHz.}  \label{fig:sodium_ThetaB}
\end{center}
\end{figure}

A simulation of the magnetic polar angle dependence for magnetic fields strengths typical for La Palma, Paranal, and Starfire Optical Range (SOR) is shown in Fig.~\ref{fig:sodium_ThetaB}. This simulation does not include any atmospheric propagation contribution nor beam or vertical integration and it has the purpose of isolating the behaviour of chirping as a function of the magnetic polar angle only. Although the specific return flux is in all cases a monotonically decreasing function with increasing $\theta_B$, the enhancement factor raises with increasing $\theta_B$ which supports the northward drift of the enhancement factor observed in the sky plots for La Palma.


\section{Conclusions}

The simulations conducted in this study suggest that a significant enhancement in the return flux from sodium LGS can be obtained with frequency chirping of the guidestar laser. The enhancement factor depends mainly on the chirp rate and on the irradiance in the sodium layer. We have developed an empirical expression for the optimal chirp rate within the irradiance regime of operation; however, in a real scenario, the distribution of laser irradiance in the mesosphere is a dynamic function of atmospheric turbulence. A real-time monitoring of atmospheric seeing at the time of laser propagation could be useful to estimate the effective LGS irradiance in order to continuously optimize the chirp rate for maximum enhancement. Ideally, wavefront correction of the uplink laser could be used to obtain a more stable and high-irradiance spot in the mesosphere suitable for optimal chirping.

Our simulations indicate no strong dependence of the optimal chirp rate on the magnetic polar angle. In fact, the return flux always decreases for increasing magnetic polar angle due to Larmor precession. Simulations suggest that for a given chirp rate there is a higher effectiveness of chirping (larger enhancement factor) with increasing magnetic polar angles. The peak enhancement as a function of $\theta_B$ also depends on the strength of the magnetic field. The physical mechanisms that explain this effect are complex as there is a dynamic interplay between the precession of polarized atoms in a resonant velocity class and the chirping through the velocity distribution which carries precessing polarized atoms into the next velocity group (due to recoil) and that also pumps fresh unpolarized atoms. The increasing efficiency of chirping with increasing $\theta_B$ can be interpreted as a counteracting effect of Larmor precession up to the point where precession is strong enough (large $\theta_B$ and/or large magnetic field) such that it lowers the efficiency of chirping again. 

Realistic simulations of the return flux for the site of La Palma, which included the influence of average atmospheric conditions in the mesosphere and also the distribution of intensities on a Gaussian beam, show a significant enhancement up to 60\% using current generation laser guide star systems. It is worthwhile performing on-sky experiments to demonstrate the benefits of chirping in real conditions, as currently ongoing using the ESO Wendelstein Laser Guide Star facility located at ORM in La Palma. The confirmation of these simulations could motivate the integration of chirping schemes in the increasing number of 20 W-class lasers being commissioned in several astronomical observatories nowadays. Certainly, more return flux will be necessary in the near future to extend routine LGS-AO operations to serve instruments working at visible wavelengths and for higher-order and extreme AO. 

\section{Acknowledgement}
We thank Martin Enderlein and Frank Lison for stimulating discussions. F.P.B. acknowledges the support of a doctoral scholarship from the Carl-Zeiss Foundation. Part of this research was conducted using the supercomputer Mogon/HIMsterII and advisory services offered by Johannes Gutenberg University Mainz (hpc.uni-mainz.de).


\bibliographystyle{aa} 
\bibliography{BibList} 



\end{document}